\documentclass[english]{iopart}
\usepackage[T1]{fontenc}
\usepackage[latin1]{inputenc}
\usepackage{babel}
\usepackage{graphics}

\makeatletter

\providecommand{\LyX}{L\kern-.1667em\lower.25em\hbox{Y}\kern-.125emX\@}

\usepackage[T1]{fontenc}
\usepackage[latin1]{inputenc}


\begin{document}

\title{Random Matrix Theory and 
 higher genus  integrability: the quantum
chiral Potts model}

\author{J.-Ch.~Anglès d'Auriac \dag, 
J.-M.~Maillard\ddag,   C.M.~Viallet\ddag\ }
\address{\ddag\  Centre de Recherches sur les Tr\`es Basses Temp\'eratures, 
BP 166, 38042 Grenoble, France}
\address{\ddag\ LPTHE, Tour 16, 1er \'etage, Bo\^\i te 126,  4 Place Jussieu, 
75252 Paris Cedex 05, France}

\begin{abstract}
 We  perform a Random Matrix Theory (RMT) analysis of the
quantum four-state chiral Potts chain for different sizes of the chain
up to size $L=8$. Our analysis gives clear evidence of a Gaussian
Orthogonal Ensemble statistics, suggesting the existence of a
 generalized  time-reversal invariance.
 Furthermore a change from the (generic) GOE distribution to
a Poisson distribution occurs when the integrability conditions are
met.  The chiral Potts model is known to correspond to a
(star-triangle) integrability associated with curves of genus higher
than zero or one.  Therefore, the RMT analysis can also be seen as a
detector of ``higher genus integrability''.
\end{abstract}
 
\vskip .2cm




\section*{Introduction}

Initially developed in the framework of nuclear physics
or atomic physics~\cite{Rosen}, Random Matrix
Theory (RMT) provides a versatile
 characterization of chaos~\cite{Gu}. Since the
 pioneering work of Wigner~\cite{Wig55} Dyson~\cite{key-8},
and Mehta~\cite{Meh91}
RMT has been applied successfully to various domains of physics.
 As a limiting case 
RMT signals the emergence of integrability, which 
shows up in the change of the generic Wignerian level spacing
distribution into Poissonian or Dirac distributions. The
 first examples of this
 appeared when considering simple harmonic oscillators (totally
rigid spectrum) or free
fermions models~\cite{FaWu70,Fel73c,Bohi}. The reduction
 to Poisson distribution reflects nothing
but the independence of the eigenvalues. This  change in the
distribution may sometimes come from  a dimensional
reduction of the model, like in  the so-called disorder
solutions~\cite{rmttransfer,criterion,GeHaLeMa87}. It
can also be found in genuinely correlated
systems, the reduction being now associated to 
Bethe Ansatz integrability~\cite{Kast,Sut70,Gaudin,LiWu72}
or Yang-Baxter integrability~\cite{Bax82} with rational or elliptic functions.
 It is natural to ask
whether this link between Poisson reduction and Yang-Baxter integrability
still holds when the solutions of the 
Yang-Baxter equations {\em are no longer parametrized
in terms of abelian varieties}. The perfect example to address this
question is the chiral Potts model for which Au-Yang et al. found
 a higher genus solution~\cite{BaPeAu88} of
the Yang-Baxter equations. These solutions appeared 
in the two-dimensional {\em classical} chiral
 Potts model on an anisotropic
square lattice (see for instance~\cite{chiral}).
As a consequence of the Yang-Baxter equations, there exists a
family of commuting transfer matrices, also commuting with some quantum
hamiltonian given below for the chiral Potts model (see (\ref{hamilt})
in the following). The RMT approach
can  be applied directly to 
analyzing the eigenvalues of 
 the transfer matrices of the two dimensional
classical models~\cite{rmttransfer}, but of course it is much simpler when applied to
quantum hamiltonians, since the 
latter\footnote[1]{The fact that the Hamiltonian does not depend on the
spectral parameters does not necessarily mean that it is blind to
the abelian or non abelian character of the integrability varieties.
This appears when one tries to build the eigenvectors of the
quantum Hamiltonian of the quantum chiral Potts chain, via Bethe
Ansatz, since the method fails for higher genus spectral curves.  }
does not depend
on the spectral parameters. It is also numerically 
much more convenient.
For the sake of simplicity we restrict ourself to the RMT
analysis of the quantum hamiltonian.

The paper is organized as follows: in section~\ref{quantumPottschain}
we recall some results about the chiral Potts model.
In section~\ref{rmtmachinery} we review how to use in practice
RMT in the context of quantum hamiltonian. In section~\ref{symmetr}
we  review some symmetries of the quantum hamiltonian of the
chiral Potts model, and discuss time-reversal invariance.
Our numerical results are
presented in section~\ref{results}, where we also discuss the 
unexpected occurrence of a GOE statistics. 

\section{The quantum chiral Potts chain.}
\label{quantumPottschain}

The Hamiltonian of the quantum chiral Potts chain first introduced
by Howes, Kadanoff and den Nijs~\cite{Howes} and 
 von Gehlen and Rittenberg~\cite{vanG}
is defined by :
 \begin{equation}
\label{hamilt}
H\equiv H_{X}+H_{ZZ}=\sum _{j}H_{jj+1}
=-\sum _{j}\sum ^{N-1}_{n=1}[\overline{\alpha _{n}}\cdot (X_{j})^{n}
+\alpha _{n}\cdot (Z_{j}Z^{\dagger }_{j+1})^{n}]
\end{equation}
 where \( X_{j}=I_{N}\otimes \cdots \otimes X\otimes \cdots \otimes I_{N} \)
and \( Z_{j}=I_{N}\otimes \cdots \otimes Z\otimes \cdots \otimes I_{N} \). 
Here   \( I_{N} \) is  a \( N\times N \) unit matrix while \( X \)
and \( Z \) are \( N\times N \) matrices,
 in  $j$-th position in the tensor product,   with entries :
 \[
Z_{j,m}=\delta _{j,m}\exp [2\pi i(j-1)/N] \quad
\mbox{ and } \quad
X_{j,m}=\delta _{j,m+1}\; \; (\textrm{mod }N)\]

 The self-dual model~\cite{Dolan} corresponds
 to \( \alpha _{n}=\overline{\alpha _{n}} \).
Conformal theory analysis of the
 3-state model can be found in~\cite{DaDeKlaMcCoMe93}.
Some spectral analysis of this model have been performed 
for the quantum self-dual model or the
 3-state model~\cite{vanG,Albert,BeMaVi92}. 

In this paper we restrict ourself to the \( N=4 \) (four-state chiral
Potts model) non self-dual case. The integrability conditions
read (see equations (33a), (33b), (33c) and (33d) in~\cite{chiral}) :
 \begin{eqnarray}
\label{integrvar}
\frac{\overline{\alpha _{2}}^{2}}{\overline{\alpha
_{1}}\overline{\alpha _{3}}} 
& = & \frac{\alpha ^{2}_{2}}{\alpha _{1}\alpha _{3}}\label{var1} \\
\frac{\overline{\alpha _{1}}^{2}+\overline{\alpha
_{3}}^{2}}{\overline{\alpha _{2}}} 
& = & \frac{\alpha ^{2}_{1}+\alpha _{3}^{2}}{\alpha _{2}}\label{var2} \\
(\alpha ^{2}_{1}-\alpha ^{2}_{3})(2\alpha ^{2}_{2}-\alpha _{1}\alpha
_{3}) 
& = & 0\label{var3} \\
(\overline{\alpha _{1}}^{2}-\overline{\alpha
_{3}}^{2})(2\overline{\alpha _{2}}^{2}
-\overline{\alpha _{1}}\overline{\alpha _{3}}) & = & 0\label{var4} 
\end{eqnarray}
 There are several simple solutions like (up to a multiplicative common
factor) : 
\begin{eqnarray}
 &  & (\alpha _{1},\, \alpha _{2},\, \alpha _{3},\,
 \overline{\alpha _{1}},\, \overline{\alpha _{2}},\, 
\overline{\alpha _{3}})\, \, =\, \, (r,\, 1,\, r,\, \pm r,\, 1,\, \pm r)\\
 &  & \qquad \hbox {or:}\quad \, \, =\, \, 
(r,\, 1,\, r,\, \pm i\cdot r,\, -1,\, -\pm i\cdot r)\nonumber 
\end{eqnarray}
 One of these solutions is a self-dual solution and the other ones
are also quite trivial. In the last two equations (\ref{var3})
and (\ref{var4}) we choose the second factor of the left-hand side
to be zero.

In order to have a real spectrum
we choose \( \alpha _{1}=\alpha _{3}^{\star } \),
\( \alpha _{2}=\alpha _{2}^{\star } \),
\( \overline{\alpha _{2}}=\overline{\alpha _{2}}^{\star } \)
and \( \overline{\alpha _{1}}=\overline{\alpha _{3}}^{\star } \)
(where the star denotes the complex conjugate) yielding a hermitian
Hamiltonian. A possible parametrization is then :
\begin{eqnarray}
\label{integCoy}
\alpha _{1} & = & \alpha _{3}^{\star } 
\,\,  = \, \sqrt{1+r}+i\sqrt{1-r}\label{para1} \, , \qquad \qquad
\alpha _{2}  =  1 \label{para2} \\
\overline{\alpha _{1}} & = & \,
\overline{\alpha _{3}}^{\star }\, =\,
 \sqrt{n^{2}+rn}+i\sqrt{n^{2}-rn}\label{para3}  \, , \qquad 
\overline{\alpha _{2}} =  n \nonumber
\end{eqnarray}
 where \( r \) and \( n \) are real and such that $\,  |r| < |n|\, $
 and $\, |r| <1$.
 The value \( n=1 \) yields  the self dual situation. Note
that we can scale \( \alpha _{1} \), \( \alpha _{2} \), \( \alpha _{3} \),
\( \overline{\alpha _{1}} \), \( \overline{\alpha _{2}} \)
 and \( \overline{\alpha _{3}} \)
by the same common factor which enables to normalize $\, \alpha _{2}
=  1$ in (\ref{para2}).

\section{The RMT machinery.}
\label{rmtmachinery}
Performing a RMT analysis means that 
one considers the spectrum of the quantum Hamiltonian,
or of the transfer matrix, as a collection of numbers, and looks for
some possibly universal statistical properties of this collection
of numbers. Indeed, neither the raw spectrum, nor the raw level
spacing distribution, have any universal property. In order to
uncover them, one has to perform some normalization
of the spectrum: the so-called \emph{unfolding} operation. This amounts
to making the \emph{local} density of eigenvalues of the spectrum
equal to unity everywhere~\cite{BrAdA96,HsAdA93,hm4,hmth}. In other
 words, one subtracts the regular
part from the integrated density of states and considers only the
fluctuations. It is believed that the unfolded
spectra of many quantum systems are very close to one of four archetypal
situations described by four 
statistical ensembles emerging from the analysis of the (real)
spectrum of random\footnote[3]{By random matrices one means that the
entries of the matrices are {\em independent Gaussian random
variables}. This is a crucial assumption. 
Of course, if the entries are not independent Gaussian random
variables, one can get all kinds of cross-overs between these
four statistical ensembles~\cite{DupuisMont}. 
} matrices~\cite{Meh91}. For  integrable
models this is the statistical ensemble of diagonal random matrices,
while for non-integrable systems this can be the Gaussian Orthogonal Ensemble
(GOE), the Gaussian Unitary Ensemble (GUE), or the Gaussian Symplectic
Ensemble (GSE),  depending on the symmetries
of the model under consideration. One-dimensional quantum systems
for which the Bethe ansatz works have a level spacing distribution
close to a Poissonian (exponential) distribution~\cite{PoZiBeMiMo93}, 
\( P(s)=\exp (-s) \),
whereas if the Bethe ansatz does not work, the level spacing distribution
can be approximated, if the hamiltonian is time-reversal invariant, either
by the Wigner surmise for the Gaussian Orthogonal Ensemble (GOE):
\begin{equation}
\label{PGOE}
P_{\textrm{GOE}}(s) \simeq \frac{\pi }{2}s\exp (-\pi s^{2}/4)
\end{equation}
or by the Gaussian Symplectic Ensemble (GSE):
\begin{equation}
\label{PGSE}
P_{\textrm{GSE}}(s) \simeq B^{3}s^{4}\exp (-Bs^{2})
\end{equation}
where 
\( B=\left( \frac{8}{3}\right) ^{2}\frac{1}{\pi }\simeq  \)2.263.
Note that GOE can also occur in a slightly more general framework
(``false'' time-reversal violation, \( A \)-adapted basis~\cite{Rob}).
When one does not have any time-reversal symmetry,
 the Gaussian Unitary Ensemble distribution should appear :
\begin{equation}
\label{PGUE}
P_{\textrm{GUE}}(s) \simeq \frac{32}{\pi ^{2}}s^{2}\exp (-4s^{2}/\pi )
\end{equation}

The three above expressions are good approximations of the exact
$P(s)$, the latter being solutions of particular Painlev\'e
equations~\cite{JMMS80,TraWi93,FoWi02a,FoWi02b,FoWi02c}.

Two-dimensional quantum spin systems were numerically shown to yield
GOE distribution~\cite{BrAdA96,MoPoBeSi93,vEGa94}.  Other statistical
properties may also be studied, like correlations between eigenvalues
(see section (\ref{brody})), but the core of the analysis will be to
compare the level spacing distribution of the unfolded spectrum with
the Poisson and the three Gaussian distributions.

\subsection{The unfolding procedure.}
\label{unfold}
The unfolding can be achieved by different 
means~\cite{hmth}. There is however no rigorous prescription
 and the best criterion
is the insensitivity of the final result to the method employed and/or
to the parameters which any unfolding method
introduces, for reasonable variation. We denote
 \( E_{i} \) the raw eigenvalues and \( \epsilon _{i} \)
the corresponding ``unfolded'' eigenvalues. The unfolding requirement
is that the local density of the \( \epsilon _{i} \)'s is equal to
one. One needs to compute an averaged integrated density of states
\( \overline{\rho }(E) \) from the actual integrated density of states:
\[
\rho (E)=\frac{1}{N}\int ^{E}_{-\infty }\sum _{i}\delta (e-E_{i})de\]
 and then we take \( \epsilon _{i}=N\overline{\rho }(E_{i}) \). In
order to compute \( \overline{\rho }(E) \) from \( \rho (E) \),
several methods are possible. One can choose a suitable odd integer
\( 2r+1 \) of the order of 9--25 and then replace each eigenvalue
\( E_{i} \) by a local average: \begin{equation}
E_{i}^{\prime }={1\over 2r+1}\sum _{j=i-r}^{i+r}E_{j}\; ,
\end{equation}
 Then \( \overline{\rho }(E) \) is approximated by the linear interpolation
between the points of coordinates \( (E_{i}^{\prime },i) \). Another
method consists in replacing each delta peak in \( \rho (E) \) by
a Gaussian distribution centered at the location of the peak
and with a properly chosen mean square deviation. There are two ways
to choose this variance: one can set the same mean square deviation
for every peak, or even better, one chooses a different mean square
deviation for each peak, so that the number of neighboring peaks inside
half-width of the Gaussian distribution is kept constant. Another
method is to discard the low frequency components in a Fourier
transform of \( \rho (E) \). A detailed explanation and tests of
these methods of unfolding are given in~\cite{BrAdA97}. Note
that all these methods require an adjustment parameter (the number
\( r \) defining the running average, the mean square deviation itself
or the number of neighboring peaks inside half width for Gaussian
unfolding, the cut-off for Fourier unfolding). When this adjustment
parameter is large, the smoothing becomes too efficient, and
the fluctuations are washed out. By contrast  too small an adjustment
parameter gives a totally rigid level spacing: the unfolded integrated
density of states coincides with the raw integrated density of states.
Out of the three methods, the moving average unfolding
is the fastest one, but the Gaussian with adapted mean square
deviation gives the best results. Notice that
 extremal eigenvalues are discarded since
they are affected by  finite size effects and this
 introduces another, slightly
less pertinent, adjustment parameter.

\subsection{Quantities characterizing the spectrum \label{brody} }

Once the spectrum has been computed, sorted and unfolded, various
statistical properties of the spectrum can be investigated. The simplest
one, which is also the most significant, and the most universal, is
the distribution \( P(s) \) of  level
 spacings \( s=\epsilon _{i+1}-\epsilon _{i} \)
between two consecutive unfolded eigenvalues $\epsilon _{i}$ 
and $\, \epsilon _{i+1}$. The distribution \( P(s) \)
will be compared to an exponential distribution and to the GOE Wigner
law (\ref{PGOE}). Usually, a simple visual inspection
is sufficient to recognize the presence of 
{\em level repulsion}~\cite{Rosen}, the main
property for non-integrable models. In order to quantify the degree
of level repulsion, it is convenient to use a parametrized distribution
which interpolates between the Poisson law and the GOE Wigner law.
>From the many possible distributions, we have chosen the \emph{Brody
distribution} : 
\begin{equation}
\label{brodis}
P_{\beta }(s)=c_{1}s^{\beta }\exp (-c_{2}s^{\beta +1})
\end{equation}
 with
 \begin{equation}
c_{2}=\, \left[ \Gamma 
\left( {\beta +2\over \beta +1}\right) \right] ^{1+\beta }
\quad \mbox {and}\quad c_{1}=(1+\beta )c_{2}\; 
\end{equation}
 This distribution turns out to be convenient since its indefinite
integral can be expressed with elementary functions. It has been widely
used in the literature. For \( \beta =0 \), this is a simple exponential
for the Poisson ensemble, and for \( \beta =1 \), one recovers the
Wigner distribution for the GOE. Minimizing the quantity:
\begin{equation}
\phi (\beta )=\int _{0}^{\infty }(P_{\beta }(s)-P(s))^{2}\, ds
\end{equation}
 yields a value of \( \beta  \) which characterize the magnitude of level
repulsion of the distribution \( P(s) \). We 
have always found \( \phi (\beta ) \)
small. When \( -0.1<\beta <0.2 \), the distribution is close to a
Poisson law, while for \( 0.5<\beta <1.2 \) the distribution is close
to the Wigner distribution.

If a distribution is close to the Wigner distribution
(resp. the Poisson law), this means that the GOE (resp. the Diagonal Matrices
Ensemble) correctly  describes  the unfolded spectrum, but only at
the level of neighboring eigenvalues. If one wants to go a step further
in the description of the spectrum (at a less universal
level), it is of interest to compute functions involving higher order
correlations as for example the spectral rigidity~\cite{Meh91}: 
\begin{equation}
\Delta _{3}(E)=\left\langle \frac{1}{E}\min _{a,b}\int _{\alpha
-E/2}^{\alpha +E/2}
{\left( N(\epsilon )-a\epsilon -b\right) ^{2}d\epsilon }\right\rangle _{\alpha }\; ,
\end{equation}
 where \( \langle \dots \rangle _{\alpha } \) denotes averaging
over the whole spectrum. This quantity measures the deviation from
equal spacing. For a totally rigid spectrum, as that of the harmonic
oscillator, one has \( \Delta _{3}^{\textrm{osc}}(E)=1/12 \), for
an integrable (Poissonian) system one has \( \Delta _{3}^{\textrm{Poi}}(E)=E/15 \),
while for the Gaussian Orthogonal Ensemble one has
 \( \Delta _{3}^{\textrm{GOE}}(E)=\, 
\frac{1}{\pi ^{2}}(\log (E)-0.0687)+{\mathcal{O}}(E^{-1}) \).
It has been found that the spectral rigidity of quantum spin systems
follows \( \Delta _{3}^{\textrm{Poi}}(E) \) in the integrable case
and \( \Delta _{3}^{{\textrm{GOE}}}(E) \) in the non-integrable case.
However, in both cases, even though \( P(s) \) is in good agreement
with RMT, deviations from RMT occur for \( \Delta _{3}(E) \) at some
system dependent point \( E^{*} \). This stems from the fact that
the rigidity \( \Delta _{3}(E) \) probes correlations beyond nearest
neighbors in contrast to \( P(s) \).

\section{Symmetry analysis.}
\label{symmetr}
Some symmetry properties of the chiral Potts model 
can be found in the literature~\cite{Pok}. We  briefly
sketch and discuss here the symmetries of the
chiral Potts Hamiltonian which we use in our analysis.

\subsection{First properties of the Hamiltonian.}
\label{first}

The hermiticity conditions of the Hamiltonian (\ref{hamilt})
are \( \alpha _{1}=\alpha ^{\star }_{3} \), 
\( \overline{\alpha _{1}}=\overline{\alpha _{3}}^{\star } \),
\( \alpha _{2} \)  and \( \overline{\alpha _{2}} \) real. They  are
compatible with the parametrization (\ref{para1}).

In this work we concentrate on the {\em four-state} case, for which the
operators \( X \) and \( Z \) read : 
\begin{equation}
\label{XZdefinition}
X=\left( \begin{array}{cccc}
0 & 1 & 0 & 0\\
0 & 0 & 1 & 0\\
0 & 0 & 0 & 1\\
1 & 0 & 0 & 0
\end{array}\right)
 \; \quad \textrm{and}\; \quad 
 Z=\left( \begin{array}{cccc}
1 & 0 & 0 & 0\\
0 & i & 0 & 0\\
0 & 0 & -1 & 0\\
0 & 0 & 0 & -i
\end{array}\right) 
\end{equation}
Note that :
\begin{equation}
\label{XZ}
X \, Z\, =\, i\cdot Z \, X, \qquad \qquad  X\, Z^{3}= \, -i \cdot Z^{3}\, X ,
\, \cdots 
\end{equation}

Let $\, p$ be the \( \, 4\times 4 \) (symmetric) matrix :
\begin{equation}
\label{Fourier}
p=\frac{1}{2}\left( \begin{array}{cccc}
1 & 1 & 1 & 1\\
1 & i & -1 & -i\\
1 & -1 & 1 & -1\\
1 & -i & -1 & i
\end{array}\right) 
\end{equation}
related to the \( \, Z_{4} \) discrete Fourier transform.
Note that \( p \) is symmetric and unitary. Let 
 \( \, R\,  \) be the 
spin reversal \( \, \sigma \, \rightarrow \, -\sigma \,  \)
(mod 4) :
\[
R=\left( \begin{array}{cccc}
1 & 0 & 0 & 0\\
0 & 0 & 0 & 1\\
0 & 0 & 1 & 0\\
0 & 1 & 0 & 0
\end{array}\right) \]

One verifies immediately that \( \, R \) is an involution (\(
\, R = R^{-1}\)), 
that \( R=p^{2} \), and that the conjugation by \( \, R \) permutes
the \( \, 4\times 4\,  \) matrices \( \, X\,  \) and \( \, X^{3} \)
on one side, and matrix \( \, Z\,  \) and \( \, Z^{3} \) on the
other side, i.e. : 
\begin{eqnarray}
\label{excha}
R\cdot X\cdot R^{-1}=  X^{3} ,\, \quad \quad \quad \quad 
R\cdot Z\cdot R^{-1} = Z^{3} 
\end{eqnarray}
 Matrices \( \, X^{2} \) and \( \, Z^{2} \) are invariant by the
conjugation by \( \, R \). We introduce 
  the \( 4^{L}\times 4^{L}\,  \) matrix
\( \, K_{R} \) which is the tensor product
of matrix \( \, R \),  \( \, L \) times :
\begin{eqnarray}
\label{tensprod}
 K_{R}\, =\, R\otimes R\otimes R\cdots R\otimes R 
\end{eqnarray}
This matrix is symmetric, real and involutive, and therefore also
unitary. One easily verifies, from  (\ref{excha}),
that \( K_{R} \) {\em commutes with} \( H \). Let
 \( U_{p}\) be the unitary matrix :
\begin{eqnarray}
\label{up}
 U_{p}\, =\, p\otimes p\otimes p\cdots p\otimes p, \qquad   K_{R}\, =\, U_{p}\cdot U_{p}^{t}=U_{p}^{2} 
\end{eqnarray}
One may perform the change of basis
 associated with this unitary matrix
\( U_{p} \) and the hamiltonian  (\ref{hamilt}) becomes :
\begin{equation}
\label{hamilt2}
H_{ZXX}\equiv H_{Z}+H_{XX}=
-\sum _{j}\sum ^{N-1}_{n=1}[\overline{\alpha _{n}}
\cdot (Z_{j})^{n}+\alpha _{n}\cdot (X^{\dagger }_{j}X_{j+1})^{n}]
\end{equation}
since $\, p\,  X \, p^{-1} \, = \, Z \, $ and 
$\, p \, Z \, p^{-1}  =  X^{\dagger} = X^{3}$.
Obviously both hamiltonians  (\ref{hamilt}) and  (\ref{hamilt2})
have the same spectrum.

Along these lines one should  recall the existence of a {\em duality} 
symmetry (see~\cite{vanG,Marcu} for duality
 for the classical models) exchanging 
the operators $\, X_{j}$ and 
 $\, Z_{j}Z^{\dagger}_{j+1}$ in (\ref{hamilt}). 
The dual hamiltonian is :
 \begin{equation}
\label{dualhamilt}
H_{dual}\equiv 
-\sum _{j}\sum ^{N-1}_{n=1}[\alpha _{n}\cdot (X_{j})^{n}
+\overline{\alpha _{n}}\cdot (Z_{j}Z^{\dagger }_{j+1})^{n}]
\end{equation}
and the duality amounts to permuting 
the $\, \alpha_n$'s and $\, \overline{\alpha
_{n}}$'s in (\ref{hamilt}).
It is also hermitian for
 \( \alpha _{1}=\alpha ^{\star }_{3} \), 
\( \overline{\alpha _{1}}=\overline{\alpha _{3}}^{\star } \),
\( \alpha _{2} \) real and \( \overline{\alpha _{2}} \) real.
If one compares the real spectrum of  (\ref{hamilt}) and (\ref{dualhamilt}) 
one finds (this has been checked for $\, L \, = \, 3, \, 4, \, 5, \,
6, \, 7, \, 8$) that they have the same 
real spectrum only on the representations
which are the most ``symmetric'' with respect to the color ($\, (c, \,
e) \, = \, (0, \, e)\, $ see below). This is reminiscent of the 
situation encountered in~\cite{seldu}.

\subsection{Representation theory.}
\label{represent}
Eigenstates with different quantum numbers are uncorrelated.
It is necessary to compare only eigenvalues of states having  the
same quantum numbers. This amounts to block-diagonalizing the
hamiltonian (see for instance page 1710 in~\cite{Rosen}),
and this is an \emph{essential requirement of the method}. Due to lattice
symmetries, as well as permutation of colors for the chiral
four-state Potts model (\ref{hamilt}), there 
exist a collection of operators \( S \), acting on the same
space as the hamiltonian \( H\), which are \emph{independent
of the parameters} \( \alpha _{i} \) and \( \overline{\alpha _{i}} \),
and commute with \( H \) : \( [H(\alpha _{i},\overline{\alpha _{i}}),S]=0 \).
The block-diagonalization
is done with the help of the character table of irreducible
representations of the symmetry group. Details can 
be found in~\cite{hmth,BrAdA97}.

In our case, that is for hamiltonian (\ref{hamilt})
or  (\ref{hamilt2}) or even (\ref{dualhamilt}),
 the analysis goes as follows. Matrix \( X \) is nothing
but the shift operator for the color. Introduce,  for a
chain  (\ref{hamilt}) of $\, L$ sites, the \( 4^{L}\times 4^{L}\,  \)
matrix: \( \, S_{X}\, =\, X\otimes X\otimes X\cdots X\otimes X \),
which shifts simultaneously all the spins by one. Using  (\ref{XZ})
one finds that \( S_{X} \) and the hamiltonian  (\ref{hamilt})
commute. This operator \( S_{X} \) generates the abelian group \( Z_{4} \).
As far as lattice symmetries are concerned, we assume periodic boundary
conditions. We also introduce the lattice shift operator
of one lattice spacing 
\( S_{latt} \). Because of the periodic boundary
conditions \( S_{latt} \) commutes
with  hamiltonian  (\ref{hamilt}).
Similarly \( S_{latt} \) generates the abelian group \( Z_{L} \).
Obviously \( S_{X} \) and \( S_{latt} \) commute and
the total symmetry group is generically the abelian \( Z_{L} \times Z_{4} \)
group. Note  that, because of their chirality, these
hamiltonians do not commute with the mirror symmetry which exchanges
site \( n \) with site \( L+1-n \). Therefore the space symmetry
group is not the dihedral symmetry 
group \( D_{L} \). However, if some additional
conditions on the parameters $\, \alpha_n$, $\, \overline{\alpha_{n}}$
are verified, the symmetry group \( D_{L} \) can reappear.

\begin{itemize}
\item The hamiltonian 
\( H_{X}\) is hermitian
iff \( \overline{\alpha _{1}}= \overline{\alpha _{3}}^{\star } \) and 
 \( \overline{\alpha _{2}}= \overline{\alpha _{2}}^{\star } \).
The lattice space symmetry group of 
\( H_{X} \) 
is always the dihedral group \( D_{L} \),
and its spin symmetry group is generically 
(i.e. for \( \overline{\alpha _{1}}\neq \overline{\alpha _{3}} \)) 
the group \( Z_{4} \),  and becomes 
 the dihedral group \( D_{4} \) when
\( \overline{\alpha _{1}}=\overline{\alpha _{3}} \).
\item The hamiltonian \( H_{ZZ} \) is hermitian iff
\( \alpha _{1}=\alpha ^{\star }_{3} \) and \( \alpha _{2} \) is
real. The lattice space symmetry group of \( H_{ZZ} \)
is generically the group \( Z_{L} \),
and is the dihedral group \( D_{L} \) when 
\( \alpha _{1}=\alpha_3 \). The
spin symmetry group 
of \( H_{ZZ} \) is always
 the dihedral group \( D_{4} \). 
\end{itemize}

For generic \( r \) and \( n \) in equations
 (\ref{para1}), the total symmetry group is \( Z_{L}\times Z_{4} \).
 We always restrict ourselves to hermitian hamiltonians.
Consequently the $\, 4 \; L\, $ 
blocks are also hermitian and they have only \emph{real}
eigenvalues. The diagonalization is performed using standard methods
of linear algebra (contained in the LAPACK
 library~\cite{lapack}). The projector
used to block diagonalize the hamiltonian are :
\begin{equation}
\label{proj}
P_{e,c}=\left( \sum ^{L-1}_{n=0}\omega
^{en}S^{n}_{\textrm{latt}}\right) 
\otimes \left( \sum ^{3}_{k=0} i^{ck} S_{X}^{k}\right) 
\end{equation}
with \( \omega =\exp (2\pi i/L). \) This formula specifies the notations
used in the rest of the paper, the representations being indexed by
\( (e,c) \) with \( 0\leq e<L \) and \( 0\leq c<4 \). 
Keep in mind that this block diagonalization (\ref{proj})
is valid for (\ref{hamilt})
or  (\ref{dualhamilt}). For (\ref{hamilt2}),
the generator \(  S_{X}\) of the \( Z_{4} \) group is
replaced by another matrix \(  S\), similar to \(  S_{X}\),
and (\ref{proj}) is modified accordingly. We will denote $\, P_{ZXX}$
this unitary transformation.

\subsection{Time-reversal invariance and beyond  : the origin of 
 GOE statistics.}
\label{timerev}

Th existence  of a  time reversal 
invariance of the Hamiltonian  changes the  generic
GUE distribution into another distribution~\cite{Wig55,key-8,Meh91,wigner}. 

The anti-unitary time-reversal operator can be written as the composition
of a unitary operator K with the complex conjugation  C  : 
\begin{equation}
\label{timerever}
T\, =\, K\cdot C
\end{equation}
In the standard case $\, K$ is a tensor product
over all sites of the chain of some spin operator
(see for instance equations (26.13b) in~\cite{wigner}).
In the following we will have to use a more general notion of
time-reversal invariance : $\, K$ will not be necessarily
a tensor product. We will only impose
 that the unitary operator $\, K$ is a {\em constant}
matrix which  should not depend on the parameters
of $\, H$. For instance, for hamiltonian (\ref{hamilt}),
 $\, K$ must be independent of the
 $\,\alpha_n$'s and $\, \overline\alpha_n$'s.

In appendix (\ref{timerev2}) it is shown that the  time-reversal 
invariance of the hamiltonian $\, H$ implies
 that \( K \) {\em must be either a symmetric
 or an antisymmetric unitary
matrix}, together with the following relation 
between the unitary operator K and the hermitian hamiltonian $\, H $:
\begin{equation}
\label{HR}
H\, =\, \, K\cdot H^{*}\cdot K^{-1}\,  =\, K\cdot H^{t}\cdot K^{-1} 
\end{equation}
or equivalently :
\begin{equation}
\label{HR2}
 K \cdot H^{t} \, = \,  H \cdot K 
\end{equation}
 where \( \, H^{*}\,  \) and \( \, H^{t}\,  \)  are, respectively,
 the complex conjugate, and the transpose, of the hermitian hamiltonian H.

\vskip .2cm

\( \bullet  \) Consider first the case 
where \( K \) is a {\em symmetric
and unitary matrix}. Any symmetric and unitary \( \, K\,  \)
can be written (see for instance page 224 of ref~\cite{Meh91}) as
the product of a unitary operator \( \, U\,  \) and its transpose,
namely \( \, \, K\, =\, U\cdot U^{t} \), and thus the time reversal symmetry
equation  (\ref{HR}) becomes 
\( U\cdot U^{t}\cdot H^{t}\, =\, H\cdot U\cdot U^{t} \)
or equivalently:
\begin{equation}
\label{basechgt}
(U^{-1} H \cdot U)^{t} = U^{-1} H \cdot U
\end{equation}
In other words, $U$ defines a change of basis bringing $H$ into a 
symmetric form  \( H_{(s)}=U^{-1} \cdot H\cdot U \).
The  Hamiltonian \( \, H \)
being hermitian,  \( H_{(s)} \)
is also hermitian and is therefore real symmetric.
Its level spacing distribution should have a Gaussian orthogonal
ensemble statistics if this real symmetric matrix is generic enough.

\( \bullet  \) Consider now the case where relation (\ref{HR2}),
is verified with an \emph{antisymmetric} unitary matrix \( \, K \).
The order of the matrix is necessarily even, namely \( \, 2N \), otherwise
the matrix is singular. One can then perform a unitary change of basis
\( \, H\, \longrightarrow \, U^{-1}H\cdot U\,  \) where \( \, U \)
not only belongs to the \( \, N \)-dimensional symplectic~\cite{key-2}
group \( \, Sp(N) \), but is quaternion real~\cite{key-3,key-4}.
In that case the level spacing distribution will have a Gaussian
Symplectic Ensemble statistics~\cite{key-8,Meh91}.

 The time-reversal symmetry is a particular
case of invariance of the hamiltonian under the action of an
anti-unitary\footnote[5]{
In this respect we can also recall the work of von Gehlen~\cite{vonG5371,nonherm}
where a $\, Z_2$-symmetry (a Lee-Yang symmetry at 
zero magnetic field)  survives for non-zero magnetic field 
as an anti-unitary symmetry on a non-hermitian Hamiltonian.} 
operator A. It can be shown~\cite{Rob}, 
that, provided the hamiltonian has a so-called ``A-adapted basis'' (which
is the case if A is an involution), the system may display
 a Gaussian orthogonal ensemble rather than the GUE, even if
the hamiltonian has neither time-reversal invariance nor geometric
symmetry.

The form of condition (\ref{HR2}) { does not depend on the
representation of the hamiltonian}.
Performing a unitary change of basis : $\, H \, \rightarrow \, H'\, =
\, U \cdot H \cdot U^{-1}$, 
one gets from (\ref{HR2}) that :
\begin{equation}
\label{chgtuni}
  H' \cdot K'  \, = \,
  K' \cdot (H')^{t}\quad \hbox{with :} \quad 
K' \, = \, \, U \cdot K \cdot U^{t}
\end{equation}
where $\, K'$ is still a symmetric unitary matrix.  
Notice that $K$ does not transform by conjugation. 

Let $\, P$ denote the change of basis which block-diagonalizes
the Hamiltonians, and  ${\alpha}$, $\, \beta$ denotes
the indices of the blocks. $\, H$ 
may be represented by the  $\, H_{\alpha}$'s and $\, K_{block} \, = \, P 
\cdot K \cdot P^{t}$
given by  its blocks $\, K_{\alpha, \beta}$'s.
Condition (\ref{HR2}) becomes :
 \begin{equation}
\label{HKredu}
 H_{\alpha} \cdot K_{\alpha,\alpha}\, = 
\, K_{\alpha,\alpha} \cdot H_{\alpha}^{t}
\quad \quad \mbox{and}\quad \quad 
 H_{\alpha} \cdot K_{\alpha,\beta}\, 
= \, K_{\alpha,\beta} \cdot H_{\beta}^{t}
\end{equation}

\vskip .3cm 

\emph{Remark :} It is crucial that 
the  unitary operator $\, K$ is a {\em constant}
matrix. Introducing 
 the unitary matrix \( V \) diagonalizing an hermitian
operator \( H \), and   $\, \Delta$  the diagonal matrix of {\em real}
eigenvalues of $\, H$, one sees that
$\, V \cdot H \cdot V^{-1} \, = \, \Delta \, = \, \Delta^{*} \, = \, 
V^{*} \cdot H^{*} \cdot (V^{-1})^{*}\, $ $= \, V^{*} \cdot H^{*} \cdot
(V^{*})^{-1}\, $ or:
\begin{equation}
H \cdot V^{-1}\cdot V^{*} \, =
 \, V^{-1}\cdot V^{*} \cdot  H^{*} \, = \,V^{-1}\cdot V^{*} \cdot  H^{t}
\end{equation}
One thus sees that the matrix $ \, V^{-1}\cdot V^{*} $, which 
is a symmetric unitary matrix,
is a solution of (\ref{HR2}). However this matrix 
 strongly depends on the parameters of
 $\, H$. From a statistical
ensemble point of view this means that the 
ensemble of hermitian matrices cannot be reduced to 
 the ensemble of real hermitian matrices 
with an independent Gaussian distribution
for the entries in each case.
The symmetric unitary matrix $\, K$ we 
are looking for, has to be independent
of the $\,\alpha_n$'s and
 $\, \overline\alpha_n$'s  parameters.

\vskip .3cm

\section{Results. }
\label{results}

One question addressed in this paper is to decide whether or not the
RMT analysis can  detect ``higher genus
integrability''. One should recall that the quantum hamiltonian (\ref{hamilt})
exhibits genus zero integrability for self-dual
 case (\( \alpha _{i}=\overline{\alpha _{i}} \)),
or free fermions integrability for some algebraic conditions.
In order to avoid these simple cases of integrability 
and stick to higher
genus integrability, we choose to move, 
in the \( \alpha _{i},\overline{\alpha _{i}} \)
parameter space, along a trajectory crossing the integrable variety
given by 
(\ref{var1}) to (\ref{var4}). In order to avoid
the self-dual case, we choose \( n\neq 1 \)
and  fix \( r \). From these values of \( r \)
and \( n \) we deduce the values of \( \alpha _{1}=\alpha ^{\star }_{3} \)
and \( \overline{\alpha _{1}}=\overline{\alpha _{3}}^{\star } \)
and \( \overline{\alpha _{2}} \) using 
the parametrization (\ref{para1}). The trajectory
 in the parameter space is obtained by varying
 \( \alpha _{2} \). In the following we will always
consider the following trajectories :
\begin{eqnarray}
\label{traj}
\alpha _{1} & = & \alpha _{3}^{\star }  = \, \sqrt{1+r}+i\sqrt{1-r}\,
, \qquad \qquad \quad  
\alpha _{2}  =  \, t \nonumber \\
\overline{\alpha _{1}} & = & \overline{\alpha _{3}}^{\star }  = \, 
 \sqrt{n^{2}+rn}+i\sqrt{n^{2}-rn} \,, \qquad \quad  
\overline{\alpha _{2}}  =  n
\end{eqnarray}
 where \( t \), \( r \) and \( n \) are real parameters.

Integrability on this trajectory appears 
at the value \( \alpha _{2}=1 \). We concentrate on the value
 of the best \( \beta _{\textrm{brody}} \)
deduced from (\ref{brodis}) as a function of the parameter 
\( t \, = \, \alpha _{2} \).

We have constructed  the quantum Hamiltonian of the
four state Potts model for various chain sizes, up to eight 
(\( L=8 \)), i.e. matrices of size up to \( 4^8 \times 4^8 \, =
\, 65536 \times 65536 \).
Since the size of the Hilbert space grows as \( 4^{L} \), it is difficult
to go much further. The results displayed below show that the size \( L=8 \)
is sufficient to answer the question we addressed. Using the complex
characters and projectors associated with the group \( Z_{L}\times Z_{4} \)
(see (\ref{proj})) we have performed 
the block diagonalization of the hamiltonian. The
dimensions of the \( 8\times 4=32 \) blocks,
are labeled by 
\( (e,c) \) which are respectively
 to the space index in (\ref{proj})
and the color index in (\ref{proj}).  The
dimensions of the \( 8\times 4=32 \) blocks are gathered in the
following table :

\vskip .3cm

\begin{tabular}{|c|c|c|c|c|c|c|c|c|}
\hline 
&
e=0&
e=1&
e=2&
e=3&
e=4&
e=5&
e=6&
e=7\\
\hline
\hline 
c=0&
2070&
2032&
2060&
2032&
2066&
2032&
2060&
2032\\
\hline 
c=1&
2048&
2048&
2048&
2048&
2048&
2048&
2048&
2048\\
\hline 
c=2&
2064&
2032&
2064&
2032&
2064&
2032&
2064&
2032\\
\hline 
c=3&
2048&
2048&
2048&
2048&
2048&
2048&
2048&
2048\\
\hline
\end{tabular}
\vskip .3cm
 
When $\, L\, $ is a prime integer, the dimensions become
simpler : all the 
blocks have the same dimensions $\, d_{all} \, = \,
(4^L-4)/4/L\, $
except the blocks of maximal symmetry with respect to the space group
$\, Z_L\, $ : $\, (e, \, c)\, = \, (0, \, c)\, $
which have  dimension $\,d_{(0, \, c)}\, = \, 1+ d_{all}$.

 {\em We first found
for each of the 32 blocks, that the eigenvalues are not degenerate
in each block, and, furthermore, these blocks are irreducible}. 
We then performed the unfolding in each block independently.
All these calculations have been checked against full diagonalization
for small sizes, as well as for special parameter sets yielding a
real dihedral symmetry group. The behavior in the various blocks
 is not significantly different. This is not totally surprising
since the dimensions  $\, d_{\alpha}$ of the various blocks are almost
equal to the average dimension $\, d_{\alpha} \simeq \,4^{L}/(4 \times
L) \, = \, 2048 $. Nevertheless the statistics is better
for larger blocks since the influence of the boundary of the spectrum
and finite size effects are smaller.
To get better statistics we 
have also averaged the results of several blocks for the same quantum chain
size \( L \). We moreover compared the four unfolding procedures,
again getting similar results. We display the results on the largest
size \( L=8 \) for the best unfolding procedure namely the Gaussian
unfolding. Figure 1 shows two level spacing distributions \( P(s) \),
for, respectively, representation (0,0) and representation (7,3), for
\( r=0.78 \), \( n=1.7 \), and \( t=1.5 \), 
which corresponds to \( \alpha _{1}=\alpha _{3}^{\star }=1.334+i\, 0.469 \),
\( \alpha _{2}=1.5 \), \( \overline{\alpha _{1}}
=\overline{\alpha _{3}}^{\star }=2.053+i\, 1.250 \)
and \( \overline{\alpha _{2}}=1.7 \). 

\begin{center}
{\centering \resizebox*{10cm}{7cm}{\rotatebox{-90}{\includegraphics{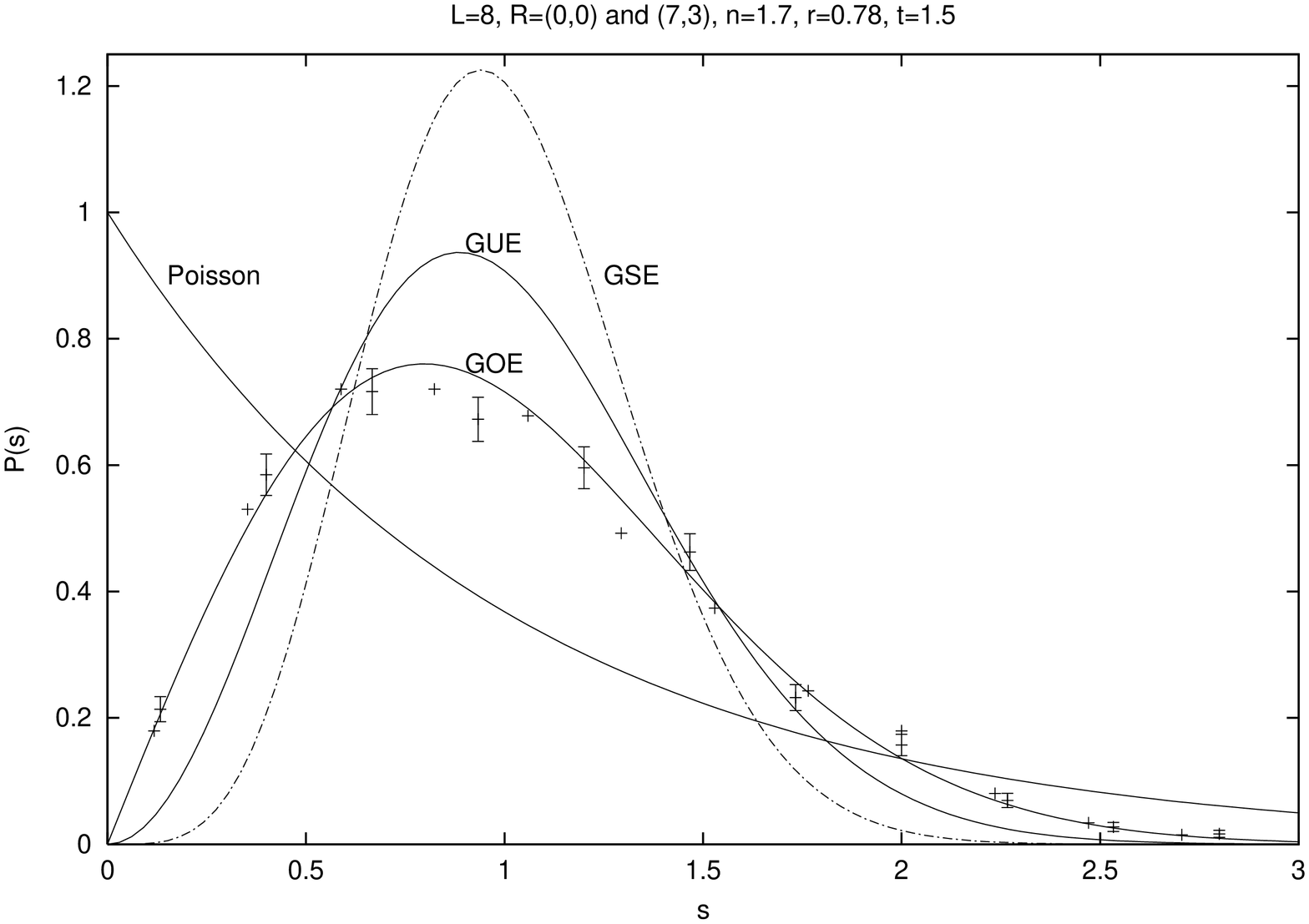}}} \par}

Figure 1: Level spacing distribution vs GOE, GUE, GSE and Poisson.
\end{center}

This figure  clearly shows  that the level spacing distribution is { close
to the GOE level spacing distribution}. The GUE
 and GSE level spacing distributions
are ruled out. Very similar results are 
obtained for the other blocks and
for others values of the parameters away from the integrability
value \( \alpha _{2}=t=1 \). We may compare 
 the Brody and the GOE  distributions 
at \( r=0.5 \), \( n=2.1 \), and
\( t=1.5 \),  corresponding  to
 \( \alpha _{1}=\alpha ^{\star }_{3}=1.225+i\, 0.707 \),
\( \alpha _{2}=t=1.5 \), \( \overline{\alpha _{1}}=\overline{\alpha
_{3}}^{\star }=2.337+i\, 1.833 \)
and \( \overline{\alpha _{2}}=2.1 \), for the block (0,0).
Figure 2 shows the level spacing distribution and the corresponding
Brody fit (\ref{brody}) for the (least square)
 best value found to be \( \beta _{brody}=0.99 \).
On the same figure the GOE level spacing distribution is also displayed,
both curves are almost indistinguishable.

\begin{center}
{\centering \resizebox*{10cm}{7cm}{\rotatebox{-90}{\includegraphics{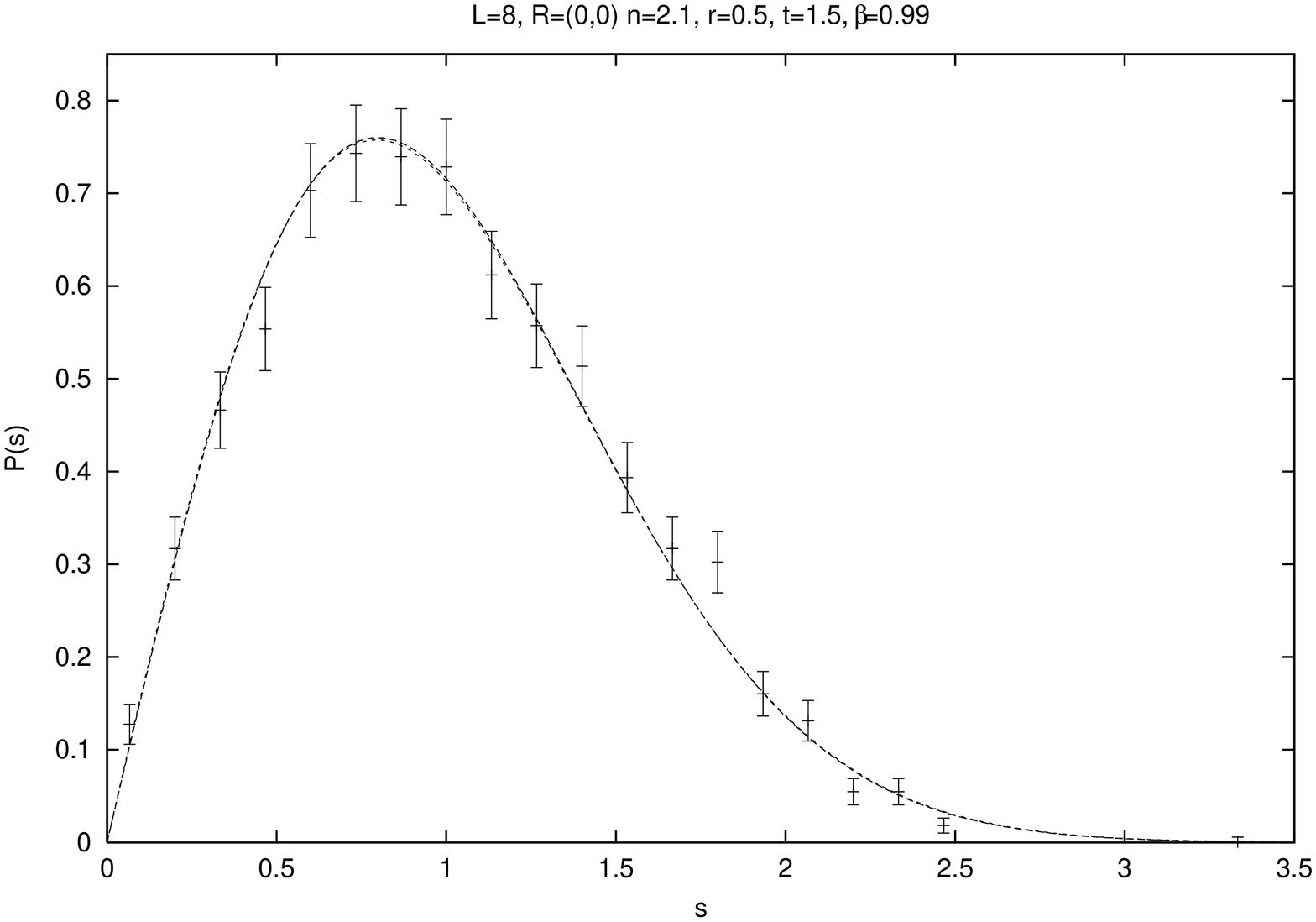}}} \par}

Figure 2: Level spacing distribution vs GOE distribution.
\end{center}

As described in section (\ref{rmtmachinery}) we can test how close
we are from the GOE statistics, considering quantities like the spectral
rigidity \( \Delta _{3}(E) \). This is depicted on figure 3 where
the spectral rigidity for the same data as in Fig. 2, and compared
with the spectral rigidity of the GOE together with the rigidity of
random diagonal matrices (Poisson).

\begin{center}
{\centering
\resizebox*{10cm}{7cm}{\rotatebox{-90}{\includegraphics{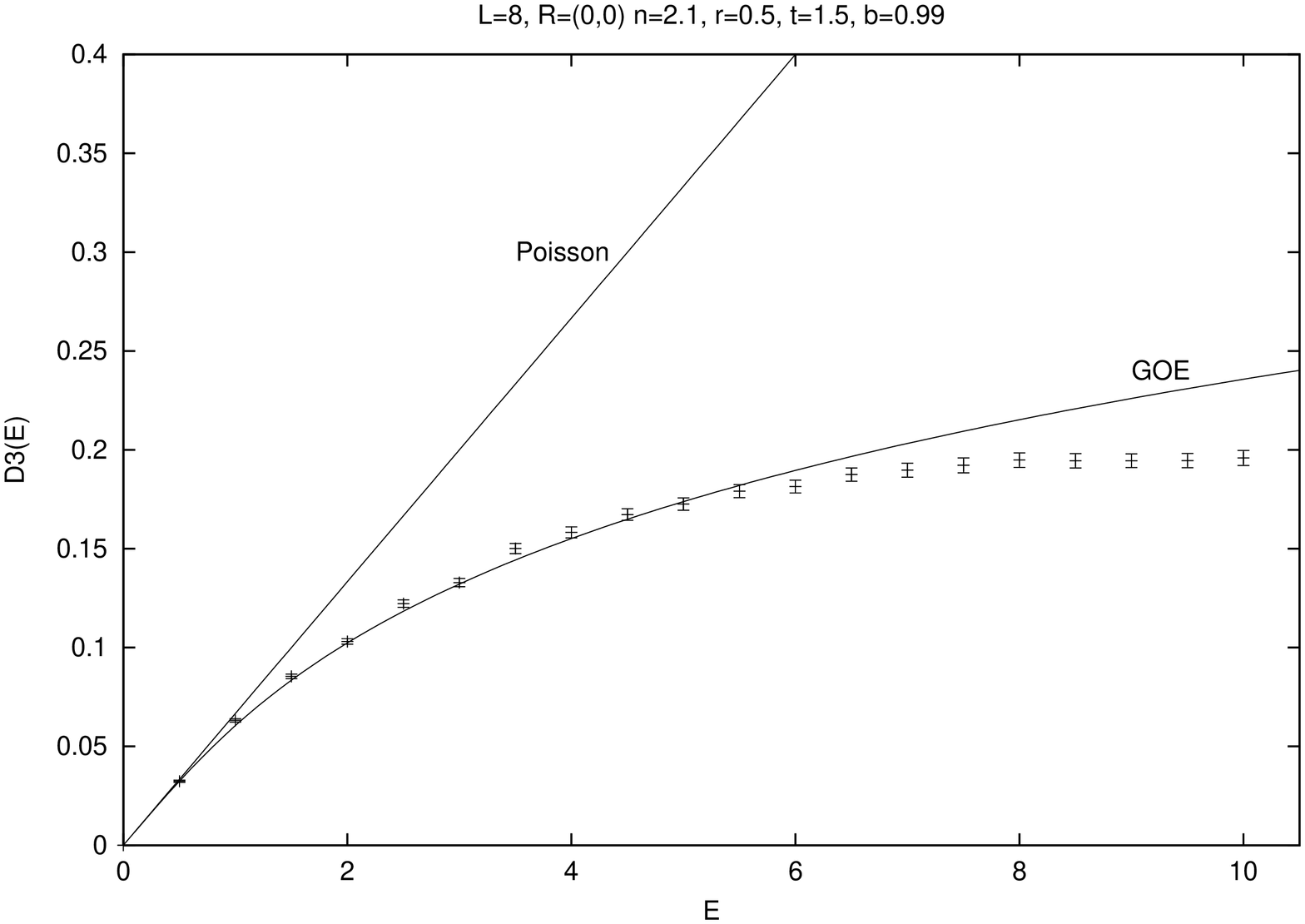}}}\par}

Figure 3: Spectral rigidity $\, \Delta_3(E)$ for $\, n=2.1$, $t=1$ 
vs spectral rigidity of the GOE.
\end{center}

The agreement with  the GOE rigidity is good up to a value of \( E\simeq 6 \),
which means that the correlation 
involving up to six consecutive eigenvalues
are properly taken into account by the GOE description.

 Figure 4 and 5 display the
level spacing distribution and the spectral rigidity for the integrable
case \( r=0.5 \), \( n=2.1 \), and \( t=1 \) which 
corresponds to \( \alpha _{1}=\alpha ^{\star }_{3}=1.225+i0.707 \),
\( \alpha _{2}=t=1 \), 
\( \overline{\alpha _{1}}=\overline{\alpha _{3}}^{\star }=2.337+i1.833 \)
and \( \overline{\alpha _{2}}=2.1 \).

\begin{center}
{\centering \resizebox*{10cm}{7cm}{\rotatebox{-90}{\includegraphics{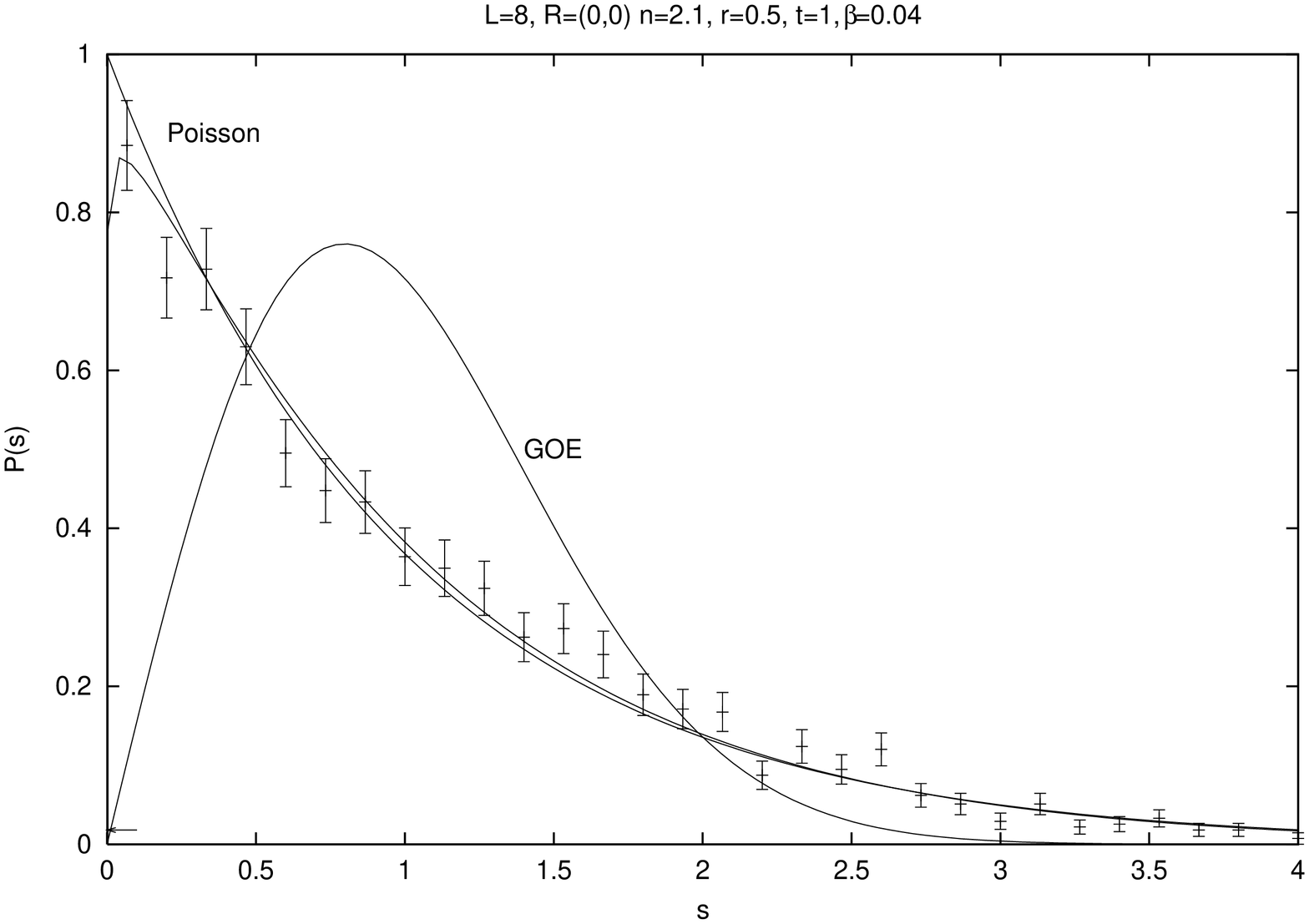}}} \par}
Figure 4: Level spacing distribution vs Poisson distribution.
\end{center}

Figure 4 shows the level spacing distribution compared to a Poisson
distribution, and also compared to the GOE level spacing distribution.
The best Brody distribution approximation of the data is found to
be \( \beta =0.04 \) using a least square fit. We have obtained very
similar results with other values of the parameters \( n \) and \( r \),
{\em and for all the} $\, 32$ {\em blocks separately}, 
when \( t \) is kept equal to
the (higher genus) integrability value \( t=1 \). This extremely 
good agreement
with a Poisson distribution is confirmed by the calculations of the
spectral rigidity displayed in figure 5. The RMT analysis can
therefore be used to { detect integrability 
even when the integrability
is not associated to abelian curves}. In other words the
independence of eigenvalues (yielding the  Poisson distribution)
is not a consequence of the abelian character of the algebraic
varieties occurring in the Yang-Baxter equations.

\begin{center}
{\centering \resizebox*{10cm}{7cm}{\rotatebox{-90}{\includegraphics{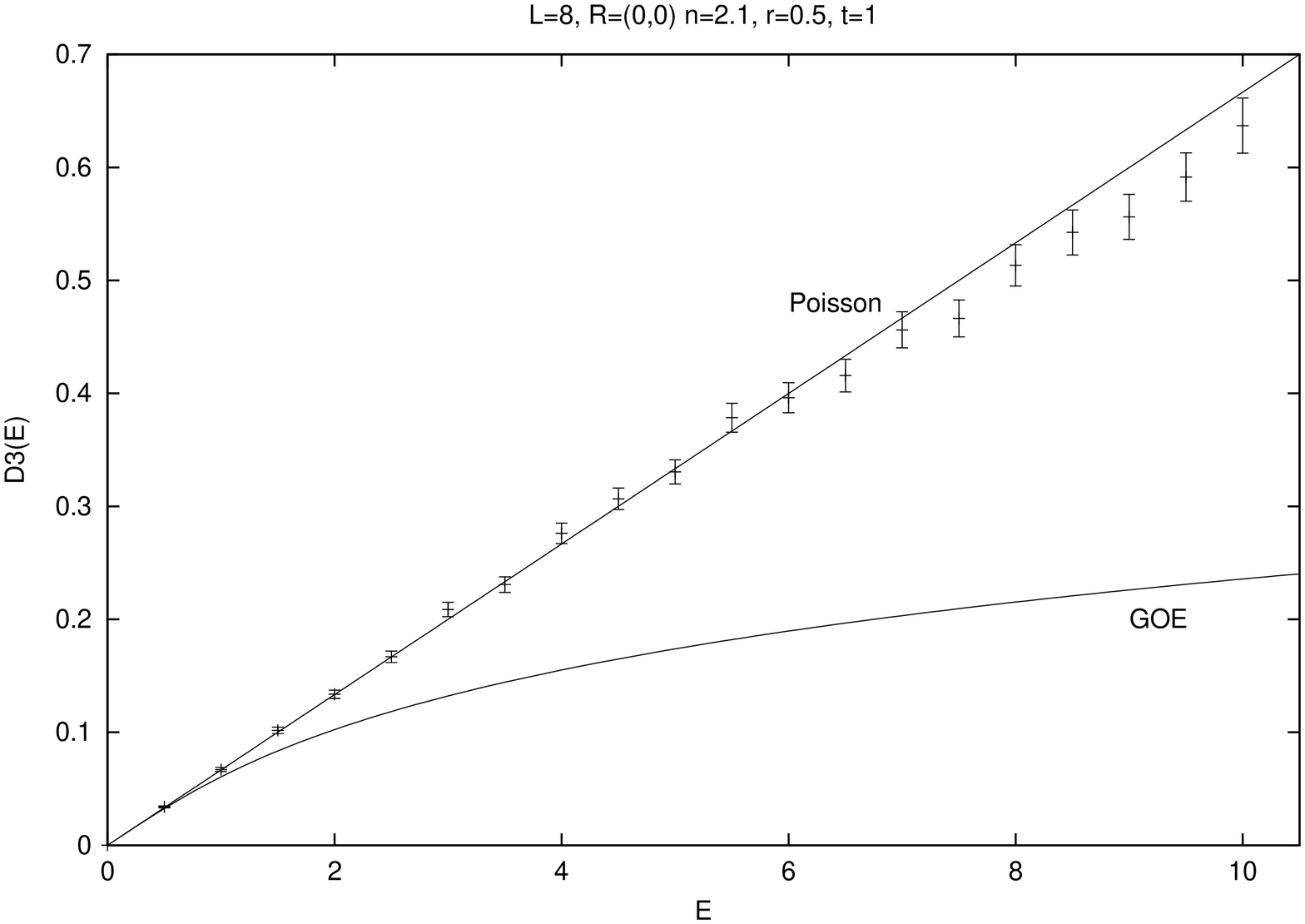}}} \par}
Figure 5:  Spectral rigidity $\, \Delta_3(E)$ for $\, n=2.1$, $r=.5$, $t=1$ 
vs Poissonian spectral rigidity.
\end{center}

This extremely good agreement with 
an {\em independent eigenvalues} situation is found
for \( t=1 \) exactly. When \( t \) is slightly different from 1,
the distribution
 is no longer Poissonian (as shown by figure 6) in agreement with the fact that
the Poissonian distribution
 should appear only at  the integrability
value \( t=1 \): as soon as \( t \) is no longer equal to 1, the
{independence} of the eigenvalues is lost, and {\em eigenvalue repulsion sets
in}. This is seen on the behavior of \( P(s) \) for small \( s \).
However, in the vicinity of \( t=1 \), the full distribution is not
exactly a Wigner distribution (\( \beta _{\textrm{brody}}=1) \),
and is an intermediate Brody distribution. We interpret this fact as
a finite size effect. Figure 6 shows the level spacing
distribution for exactly the same parameters as in figure 5 ($L=8$, $\,
r=.5$, $\, n=2.1$, for the block 
$\, R=(0,0)$) except parameter $\, t$ which is changed into \( t=1.05 \).

\begin{center}
{\centering \resizebox*{10cm}{7cm}{\rotatebox{-90}{\includegraphics{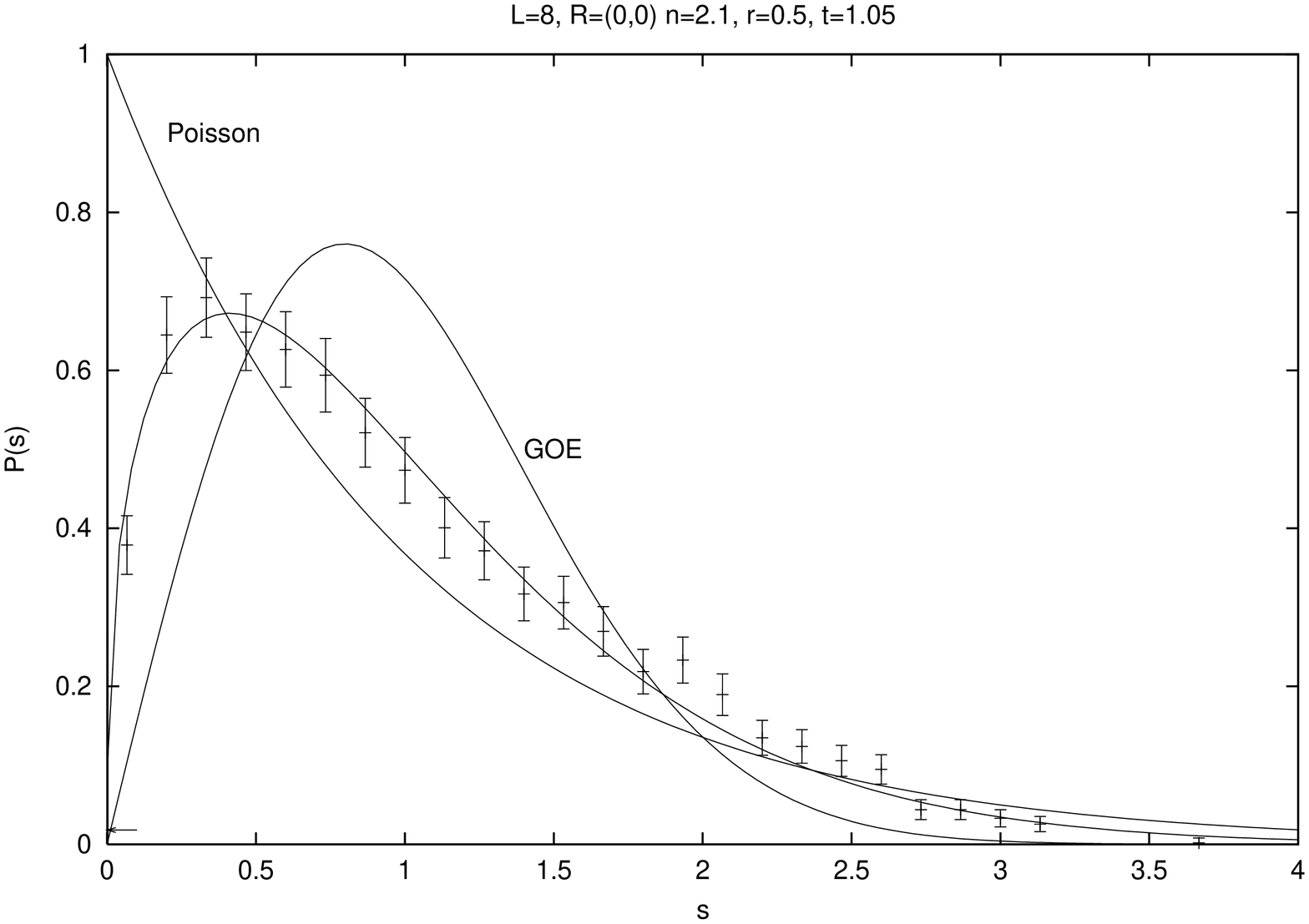}}} \par}
Figure 6:  Level spacing distribution for $n=2.1$, $r=.5$, $t=1.05$
vs GOE distribution
together with a Brody distribution for $\, \beta=.36$. 
\end{center}

The best (least square) 
fitting parameter \( \beta_{brody} \) is \( \beta_{brody}=0.36 \).
This intermediate value, between $0$ and $1$, is a consequence of the
finite size of the quantum chain. One can certainly expect 
\( \beta_{brody} \) would tend, in the thermodynamic limit,
to the GOE value \( \beta_{brody}\, = \, 1 \).
In order to quantify this (finite size) transition from integrability to chaos,
we calculate the best Brody parameter, as a function of the parameter
\( t \), keeping \( r \) and \( n \) constant. Figure 7 displays
\( \beta_{brody} \), as a function of \( t \), for all the
representations.

\begin{center}
{\centering \resizebox*{10cm}{7cm}{\rotatebox{-90}{\includegraphics{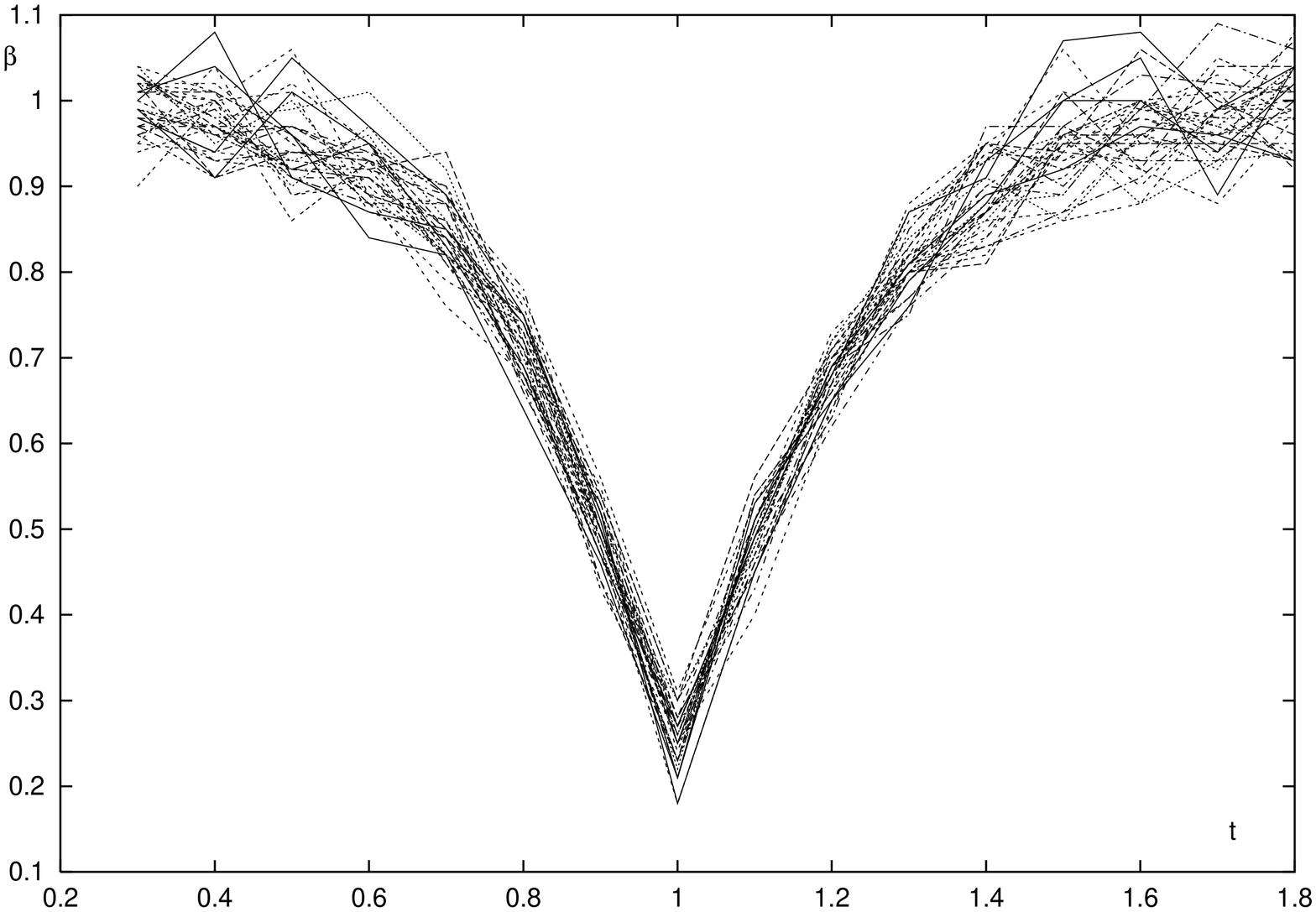}}} \par}
Figure 7:  Best $\, \beta_{brody}$ parameter as a function of
parameter $\, t$ for all the 32 representations.
\end{center}

These results confirm a  sharp transition from 
a GOE distribution to a Poisson distribution. In the thermodynamic
limit one can expect \( \beta_{brody} \) to be equal to the
GOE value \( \beta_{brody} =1\) for every value of the
parameter $\,t$, except at point $\, t=1$, where 
 \( \beta_{brody} =0\).

To make this change of regime more intuitive, we also 
show, on figure 8, a window on 
the unfolded spectrum, as a function of parameter
\( t \), for \( L=7 \). Only twenty five unfolded eigenvalues are represented.
 One sees clearly, on the unfolded spectrum, the level repulsion 
for $\, t \, \ne \, 1$ and the level repulsion weakening
around \( t=1 \).

\begin{center}
{\centering \resizebox*{10cm}{7cm}{\rotatebox{-90}{\includegraphics{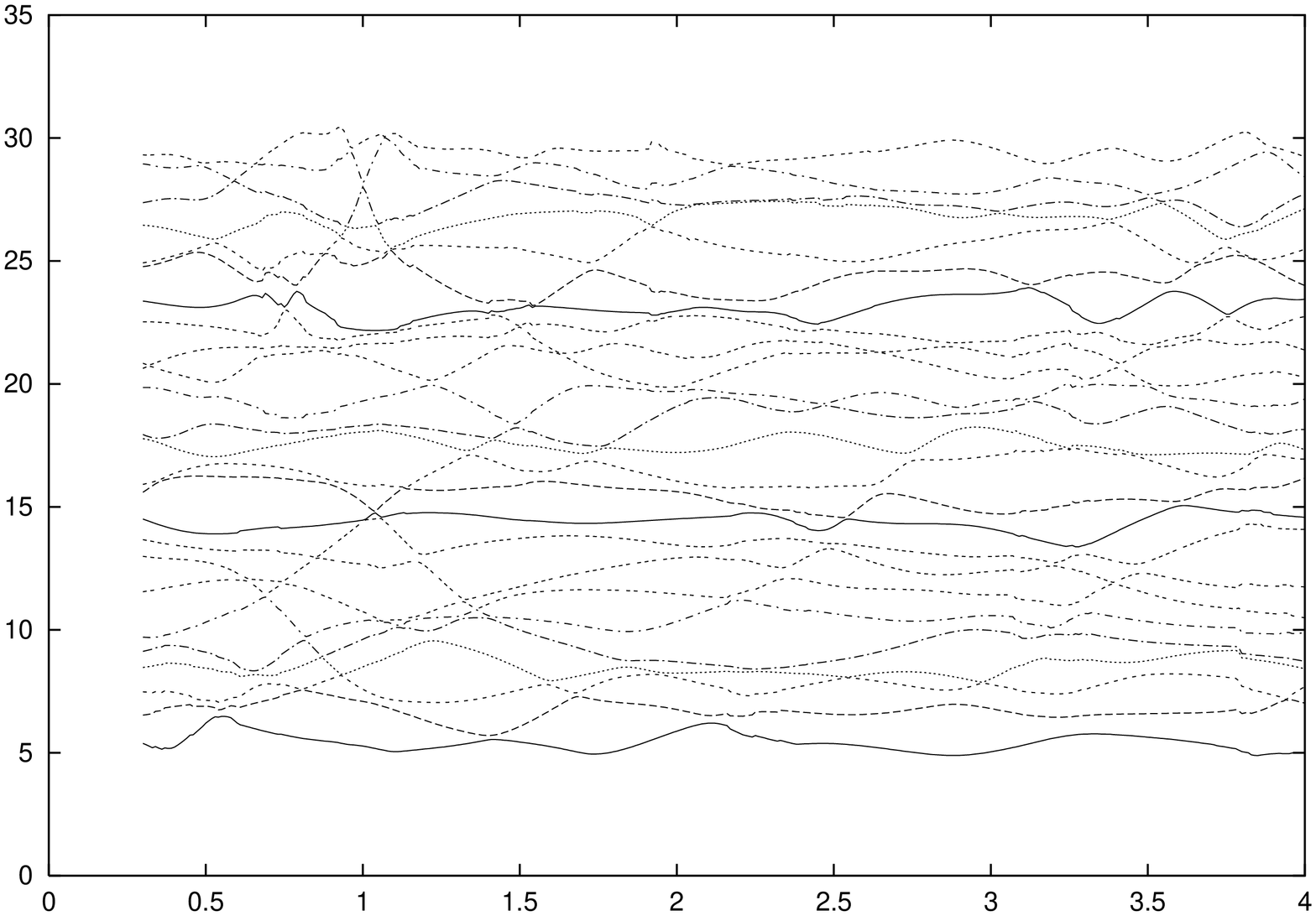}}} \par}
Figure 8: Window of the unfolded spectrum as a function of
parameter $\, t$ for seven sites.Twenty five unfolded eigenvalues
among  $\, 4^7\, = 16384$ are given as function of  $\, t$.
\end{center}

\subsection{Discussion of the occurrence of GOE.}
\label{discu}

The results presented above indicate a clear occurrence of a GOE
distribution when $\, t \ne 1$. Numerically the
 previous results  hold {\em for each of the 
$\, 4 \; L\, $  blocks}.  This certainly
requires hamiltonian (\ref{hamilt}) to have additional symmetry properties
compared to a generic hermitian matrix.
 In the following we will see that a
generalization of the  time-reversal
invariance property for the quantum hamiltonian  (\ref{hamilt}),
namely condition (\ref{HR2}) of subsection (\ref{timerev}), 
seems to hold. 

One can look for the  matrix $\, K$  of (\ref{HR2}) 
in any basis, keeping in mind the particular transformation rule
(\ref{chgtuni}). If we examine  form (\ref{hamilt2}) :
\begin{eqnarray}
\label{HZXXsymantisym}
&&H_{ZXX} \,\, = \,\, H_{Z} + H_{XX}\,\, 
=\, \,(H_{Z}\, + H_{XX}^{(s)})\, +\,\alpha_{im} \cdot H_{XX}^{(as)}  \\
&&= \, \, H_{ZXX}^{(s)}\, +\,\alpha_{im} \cdot
H_{XX}^{(as)} 
\quad \hbox{with :} \quad
\alpha_{im} \,= \,{{1} \over {2}} (\alpha _{1}-\alpha_{1}^{*})
 = \,  \,{{1} \over {2}} \,(\alpha _{1}-\alpha _{3})\, \qquad  \nonumber
\end{eqnarray}
where $\, H_{XX}^{(s)}$ is a {\em real symmetric} matrix : 
\begin{equation}
\label{hsymXX}
H_{XX}^{(s)} =-\sum _{j}[{{1} \over {2}} \, (\alpha _{1}+\alpha _{3})
\cdot (X_{j}\, X^{\dagger }_{j+1}+\, X^{\dagger }_{j}\, X_{j+1})]\, -\sum
_{j}\alpha _{2}\cdot (X_{j}X^{\dagger }_{j+1})^{2} \qquad 
\end{equation}
and $\, H_{XX}^{(as)}$ is the {\em antisymmetric}
matrix:
\begin{equation}
\label{hantisymXX}
H_{XX}^{(as)} \, = \, -\sum _{j}[
X_{j}\, X^{\dagger }_{j+1}-X^{\dagger }_{j}\, X_{j+1}]\,
\end{equation}

As a consequence of the hermiticity conditions, in particular
\( \overline{\alpha _{1}}=\overline{\alpha _{3}}^{\star } \)
with $\overline{\alpha _{2}}$ real, $\, H_{Z}$ is a real diagonal
matrix. Thus, since $\, H_{XX}^{(s)}$ is  { real symmetric},
$\, H_{ZXX}^{(s)}$ is also  { real symmetric}. In 
other words, hamiltonian 
(\ref{hamilt2}) is  { real symmetric} when  $\,\alpha_{im} \, = \, 0$,
in which case $\, K\, = K_R$ given by (\ref{tensprod}).
{\em It is thus not surprising to see the occurrence of a GOE level spacing
distribution on hamiltonian} (\ref{hamilt}) {\em when} $\,\alpha_{1}\, = 
\, \alpha_{3}\, $ {\em is real} (namely \( r=1 \) with parametrization 
(\ref{para1})).   

When $\,\alpha_{im} \, \ne  \,0$,
one looks for a matrix $\, K$, { independent of the} $\alpha_n$'s
and $\overline\alpha_n$'s, which  { commutes} with 
$\, H_{ZXX}^{(s)}\,$ and { anticommutes} with
 $\, H_{XX}^{(as)}$.

 The existence of $\, K$ implies that the non zero eigenvalues of $\,
 H_{XX}^{(as)}\, $ appear in opposite pairs,
 which we have checked up to size $\, L\, = \, 8$.

\subsubsection{Small $\, L$ cases.}
\label{L3}

$\bullet \, $ For $\, L\, = \, 3$,
the symmetric unitary 
matrices $\, K$ satisfying (\ref{HR2}) 
are not  simple  tensor products\footnote[6]{Seeking for 
matrix $\, K \, = \, M \otimes M \otimes M$ satisfying (\ref{HR2})
when $\,\alpha_{im} \, \ne \, 0$, one finds, from the commutation of $\, K$ with 
$\, H_{Z}$, that  $\,M$ must be a symmetric matrix, and from the 
anticommutation of  $\, K$ with $\, H_{XX}^{(as)}$ that 
the only solution is the null matrix. Of course when $\,\alpha_{im} \,
= \, 0$ one gets solution (\ref{tensprod}) taken for $\, L\, =3$.
},   suggesting that 
we are not in a strict time-reversal invariance 
framework (see for instance equation (26.15),
page 331 in~\cite{wigner}). For $\, L\, = 3$, 
 the $\, 64 \times 64\, $ matrices\footnote[7]{If 
\( \, K_1 \) and \( \, K_2 \) are
two unitary solutions of (\ref{HR2}), 
\( \, \, K_2\cdot K_1^{-1}\,  \) commutes with
the family of \( \, H\)'s and \( \, \, K_2^{-1} \cdot K_1\,  \) commutes with
the family of \( \, H^{t}\)'s. Thus the set of solutions of (\ref{HR2})
is related to the set of matrices commuting with \( \, H \).}
 $\, K$ satisfying (\ref{HR2})
 are linear combination of twelve 
quite simple involutive permutation matrices
with entries equal to $\, 0$ or $\,1$. For
 periodic boundary conditions,
 none of these linear combinations commute with $\, S_{latt}$, 
the lattice shift operator of one 
lattice shift spacing.

This non-trivial form of $\, K$ is confirmed
by its expression in
 the basis which block diagonalizes the Hamiltonian:
 the off-diagonal blocks
 $\,  K_{\alpha,\beta}$, $\,\alpha\, \ne \, \beta $, of
 $\, K_{block} \, = \, P \cdot K \cdot
P^{t}$ in (\ref{HKredu}) { vanish}
and { one can  restrict 
condition} (\ref{HR2})
to each block  $\, \alpha=(e, \, c)$, namely :
 \begin{equation}
\label{HKredu2}
H_{\alpha} \cdot K_{\alpha,\alpha}\, = 
\, K_{\alpha,\alpha} \cdot H_{\alpha}^{t}
\end{equation}
The off-diagonal blocks
 $\,  K_{\alpha,\beta}$ also vanish
 for hamiltonian (\ref{hamilt2}), the unitary 
transformation (\ref{proj}) being replaced by 
$\, P_{ZXX}$.

For $\, L\, = \, 3$ the symmetry group is $\, Z_4 \times Z_3$,
and there are $\, 4 \times 3 \, = \, 12$ blocks 
$\, \alpha \, = \, (e, \, c)$, labelled in short by an index 
 $\, 0, \, 1, \cdots
11$. The blocks $\, K_{\alpha,\alpha}$'s  can be written as :
\begin{equation}
\label{bloblo}
 K_{\alpha,\alpha} \, = \, \lambda_{\alpha} \cdot
k_{\alpha,\alpha}
\end{equation}
 where the $\, k_{\alpha,\alpha}$
are simple symmetric unitary matrices with as many entries as possible
normalized to $\, 1$, and where the $\, \lambda_{\alpha}$'s
are complex numbers of unit modulus.
The block matrices  $\, k_{\alpha,\alpha}$
are given in appendix (\ref{asablock}) for hamiltonian (\ref{hamilt}) 
or equivalently (\ref{hamilt2}). For instance the  block 
corresponding to the ``most symmetric'' representation, namely 
$\, \alpha \, = \, (0, \, 0)$ reads  for hamiltonian (\ref{hamilt}) 
{\em as well as for} (\ref{hamilt2}): 
\begin{equation}
\label{k00}
k_{0,0}\, = \, \,  \,  \, 
\left [\begin {array}{cccccc} 
1&0&0&0&0&0\\
\noalign{\medskip}0&0&0&1&0&0\\
\noalign{\medskip}0&0&1&0&0&0\\
\noalign{\medskip}0&1&0&0&0&0\\
\noalign{\medskip}0&0&0&0&1&0\\
\noalign{\medskip}0&0&0&0&0&1
\end {array}\right ]\,
\end{equation}
It is important to note that,
up to the multiplicative  (unit modulus) factors
 $\, \lambda_{\alpha}$'s, the
blocks  $\, k_{\alpha,\alpha}\, $
 given in appendix (\ref{asablock}) 
are unique. This means 
that in each block $\, \alpha\, $ there 
exists {\em no non trivial
symmetry operator commuting with the family}
 (\ref{hamilt}) {\em of hamiltonians H}.
\begin{equation}
[S_{\alpha}, \, H_{\alpha}] \, = \, \, 0, \, \, \, \forall \, \alpha_{n} , \,
\, \overline{\alpha _{n}}\, , \, \, n \, = \, 1, 2, 3 \quad
\Rightarrow \, \, S_{\alpha} \, = \, Identity
\end{equation}

$\, \bullet $ For $\, L\, = \, 4$ we get 
 similar results for Hamiltonian (\ref{hamilt2}). 
  The most symmetric 
block $\, (0, \, 0)\, $ yields a $\, 20 \times 20$
involutive  permutation matrix $\, k_{0,0}$.
It is important to note that all  blocks
 are quite similar to the 
ones described in Appendix (\ref{asablock}) and are {\em unique
 up to multiplicative complex of unit modulus}.   
For $\, L\, = \, 4$,  $\, K$ 
is a $\, 256 \times 256\, $  symmetric matrix. 
If one does not impose the unitarity condition, 
the set of all  solutions 
of (\ref{HR2}) reads :
\begin{eqnarray}
\label{Lequal4}
K \, = \, \, \, \,\sum_{n=0}^{n=15} p_n \cdot A_n 
\end{eqnarray}
where the $\, A_n$'s are $\, 256 \times 256\, $
symmetric matrices whose entries are equal to $\,0$ except on 
at most one entry equal to one
for each row or column. In contrast with the $\, L\, =3$ case 
the $\, A_n$'s are singular matrices ($det(A_n) \, = 0$), they are
 not permutation matrices. For certain choice of the 
parameters $\, p_n$ 
one gets, from (\ref{Lequal4}),  a matrix $\, K$ 
which is an symmetric real matrix with entries
$\, 0$ or $\, 1$, representing an {\em involutive permutation} $\, I_1$.

$\bullet $ Similar exact calculations of the blocks of  matrix 
$\,P_{ZXX} \cdot K_{ZXX} \cdot P_{ZXX}^{t}\, $
 have been performed
for $\, L=\, 5$ and $\, L\, = \, 6$. Again one finds that the
$\, \alpha \, \ne \, \beta\, $ 
off-diagonal blocks $\,  K_{\alpha,\beta}$ vanish and that
the $\, 4 \; L\, $ blocks $\,  K_{\alpha,\alpha}$ are {\em unique
 up to multiplicative complex of unit modulus}.   

 As far as the $\, (0, \, 0)$ block is concerned 
one also finds that $\, k_{0,0}\, $  
are ($52  \times 52\,  $ for $\, L = \, 5$ 
and  $\, 178 \times  178\, $ for $\, L = \, 6$) 
simple involutive permutation matrices. 

All these results are  detailed
on a website~\cite{web}
where the various blocks $\, k_{\alpha,\beta}\, $ 
are written for $\, L\, = \, 3$, $\, L\, = \, 4$, 
the blocks such that all the entries are $\, 0$ or $\, 1$ 
(but no root of unity)
are given for  $\, L\, = \, 5$, and $\, L\, = 6$
and furthermore the full $\, 64 \times 64$ and $\, 256 \times 256\, $ 
$\, K$ matrices are written for
$\, L\, = \, 3$ and $\, 4$.

For these values of $\, L$ ($\, L\, = 3, 4, 5, 6$), 
the block matrices  $\, k_{\alpha,\alpha}$
are remarkable matrices with entries $\, 0$, or $\, 1$, or $\, m$-th
{\em root of unity} ($m\, =\, 4 \, L$).

\subsubsection{Conjecture}

It is difficult to describe { all} the $\, 4 \; L\, $ blocks $\,
K_{\alpha, \alpha}$ ($\alpha \ne \, (0, \, 0)$).  It might be easier
to describe the $\, 4^L \times 4^L$ matrices $\, K$ satisfying
(\ref{HR2}) without imposing the unitarity condition in a first step.

We conjecture that the solutions of (\ref{HR2}) are linear
combinations of $\, 4 \; L\,$ solutions which are involutive
permutations $\, I_n$, {\em in the original basis} where $\, X$ and
$\, Z$ are given by (\ref{XZdefinition}).

We conjecture moreover that the block diagonalization of 
$\, H$ leads simultaneously to a block diagonalization of $\, K$ 
into  $\, 4 \; L\,$ blocks. The unitarity condition on $\, K$
translates into a simple condition 
on a multiplicative factor for each block
(modulus equal to one condition). The choice of these factors is the
only indeterminacy.

\subsubsection{Large $\, \alpha_{im}$ limit : deformation of a
quantized spectrum still yielding GOE.}

It is difficult
to find a simple closed expression for matrix $\, K$ 
satisfying  (\ref{HR2}),  for arbitrary  $\, L$ and 
$\, \alpha_{im}$. We may examine the part 
$\, H_{XX}^{(as)}$ of the Hamiltonian and
compare with the results of level spacing 
analysis of Hamiltonian (\ref{hamilt})
 or (\ref{hamilt2}) for large $\, \alpha_{im}$.

$\bullet$  Matrix  $\, H_{XX}^{(as)}$ vanishes  except on a 
set of rows and columns where its entries are equal to $\, 0$ or 
$\, \pm 1$. Unfortunately the subspace corresponding to these 
rows and columns is not an invariant subspace of $\, H_{ZXX}^{(s)}$:
  $\, H_{XX}^{(as)}$ and  $\, H_{ZXX}^{(s)}\,$ {\em do not commute}.
Furthermore this subspace becomes ``quite large'' with increasing
chain size  $L$. 
Let us consider the dimension $\, d(L)$ of this 
 subspace as a function of $\,L$. If $\, d(L)/4^L\,\rightarrow \, 0 $ when
$\, L \, \rightarrow \, \infty$ one could think that 
condition (\ref{HR2}) ``tends to be verified'' 
in the thermodynamic limit. In fact this is not the case : 
the ratio of  $\, d(L)/4^L\,$ 
is a monotonic increasing function of $\, L$.
For $L$ running from $\, L=3$ to $12$, one gets the following values,
for successive ratio of  $\, d(L)/4^L\,$:
 $\,.375, \,.406, \,$ ...,  $ .647, \,$ $ .663, \,.677$.

 More specifically, the antisymmetric 
matrix $\, H_{XX}^{(as)}$
has the same eigenvalues as 
the diagonal matrix $ \,H_{ZZ}^{im}$ :
\begin{equation}
\label{hantisym}
H_{ZZ}^{im} \, = \, -\sum _{j}[
Z_{j}\, Z^{\dagger }_{j+1}-Z^{\dagger }_{j}\, Z_{j+1}]\,
\end{equation}
that is to say the 
relative integers 
$\, 0$, $\, \pm 4$, $\, \pm 8$, $\, \pm 12, \cdots \pm 4\, m$.
 For instance, for $\, L=12$, one 
gets the eigenvalues $\, \pm 4, \, \pm 8, \,
\pm 12, \, \pm 16,\, \pm 20, \,\pm 24\, $  
respectively $\, 3920928,$ $ \,\,1471932 , $
$\,\,268752 , $$\,\,21384 , $
$\,\,528, $$\,\, 4  $ times.

$\, \bullet \, $ We may  go back to the level spacing distributions
and rigidity calculations detailed in section (\ref{results}),
when  $\, \alpha_{im}\,$ is large. This is an interesting 
situation where the spectrum of eigenvalues should be (up to the
 multiplicative factor $\, \alpha_{im}$)
a deformation of a  set
of integers $\, 0$, $\, \pm 4$, $\, \pm 8$, $\, \pm 12, \cdots \pm 4\, m$.
Let us consider again $\, L=8$ but for a large enough value of
 $\, \alpha_{im}\, = \, 200$.  
Figure 9 shows the integrated density of eigenvalues 
for $\, L\, =\, 8$. It is clear that
the eigenvalues are  (up to a multiplicative
factor $\, \alpha_{im}$) mainly located around 
the set of relative integers  
$\, 0$, $\, \pm 4$, $\, \pm 8$.
\begin{center}
{\centering
\resizebox*{10cm}{7cm}{\rotatebox{-90}{\includegraphics{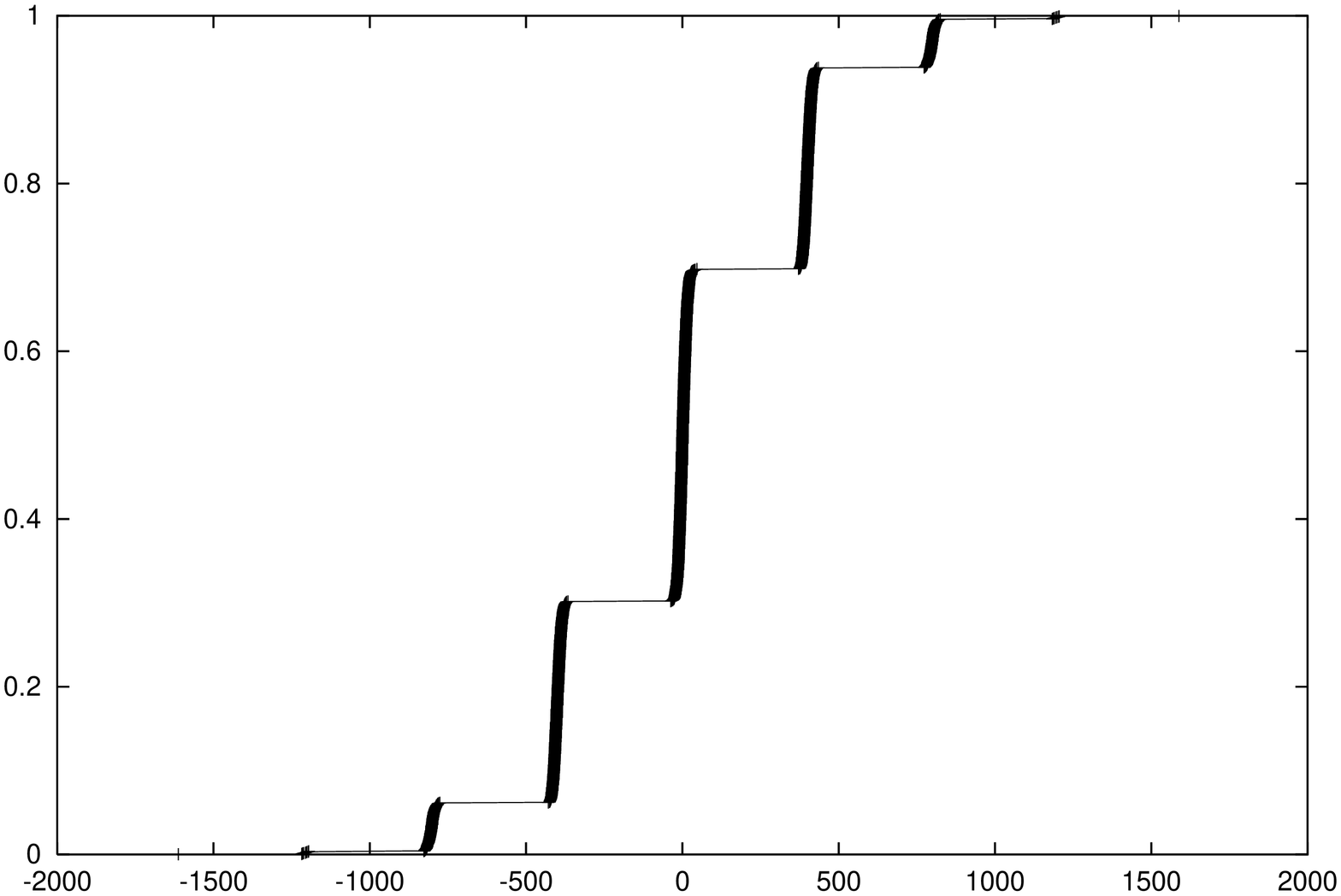}}} \par}
Figure 9: Integrated density of eigenvalues 
\end{center}

Figure 10 shows the corresponding level spacing distribution
compared to the Poisson distribution, 
to the GOE distribution, and to the best (least square) fit by a
Brody distribution. The agreement with a GOE statistics
 is extremely good since one gets
$\, \beta_{brody} \simeq .93$ :
\begin{center}
{\centering
\resizebox*{10cm}{7cm}{\rotatebox{-90}{\includegraphics{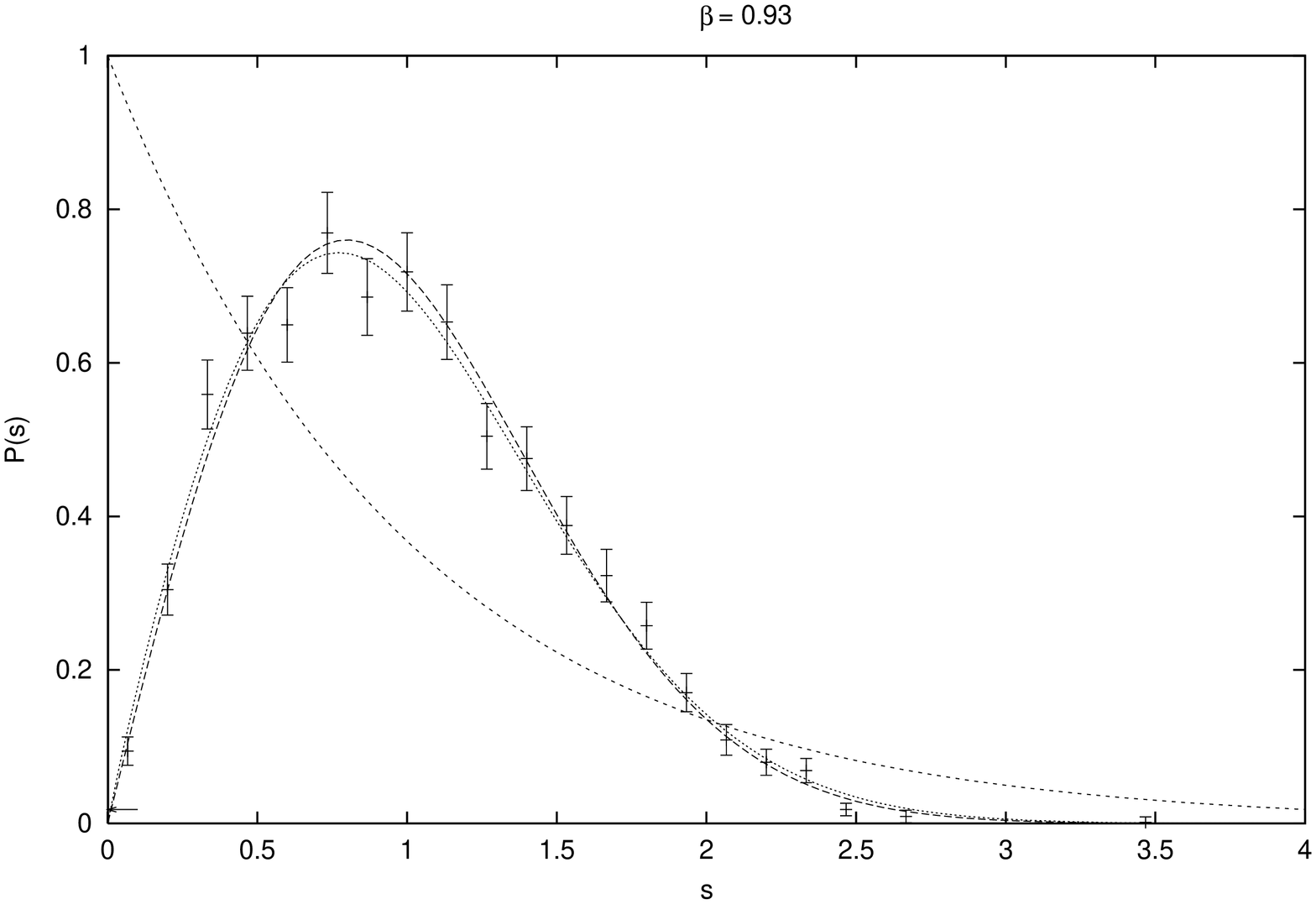}}} \par}
Figure 10: Level spacing distribution vs Poisson, GOE
distributions and Brody distribution for $\beta_{brody}=.93$
\end{center}
Recalling the analysis which yields 
figures 3 and 5 in section (\ref{results}), one also can perform rigidity 
calculations in this strong $\, \alpha_{im}$ limit. Similarly to
the results displayed in figure 3, the rigidity analysis 
 confirms, for each of the 32 representations, this GOE distribution. 
This is a non trivial limit. This strong $\, \alpha_{im}$ limit
yields a spectrum which is a {\em deformation of a spectrum
of relative integers}. The unfolding procedure yields 
a level spacing distribution which is
 still GOE! It does not matter that
the eigenvalues are concentrated near integers: what matters is the 
distribution of eigenvalues around these integers which still yields
the universal GOE level spacing distribution.
 
This very good agreement
is a strong indication of the GOE character of the level spacing
distribution of the hermitian hamiltonian (\ref{hamilt}) in general.

\subsection{Strategy for finding new  integrable lattice models.}
\label{strategy}

One may use  RMT analysis
to find new integrable lattice models, which is
extremely difficult analytically especially if they are associated
to higher genus solutions of the Yang-Baxter equations.

It has been emphasized~\cite{BeMaVi92,Ma86,22}
that this type of integrability appears when the parameters verify
very specific algebraic conditions : these conditions express
that an infinite set of birational transformations degenerates into
a  {\em finite
set}~\cite{Ma86,JaMa85,MaRo94,MeAnMaRo94,HaMa88,MaRa83}. 
We thus have a constructive way to find
 new possible integrability conditions~\cite{Lets}. 
However verifying that a particular subvariety 
of the parameter space of the model allows the Yang-Baxter
(or the generalized star-triangle) relations to be satisfied,
remains a very involved analytical task. The RMT analysis 
provides us with a numerically efficient way
to verify if these algebraic 
subvarieties yield actual integrability conditions.

One can show 
that  the general four-state classical two-dimensional chiral Potts
model has a canonical elliptic parameterization.
>From this parameterization one may write down explicitly
the equations of these algebraic varieties, which
 are the only possible locations for
the higher genus integrability conditions~\cite{Prepa}. 
 Various analysis, similar to the one
summarized on Figure 7, 
of $\,\beta_{brody}$ as a function of parameter $ \, t$, on
 various trajectories (\ref{integCoy}) in the parameter
space of the quantum hamiltonian (\ref{hamilt}),  indicate that 
the integrable variety (\ref{integrvar}) is the only
 one  with higher genus. Of
course one cannot exclude the existence of higher codimension
 integrable varieties avoiding the trajectories (\ref{integCoy})
we have considered. 

\section{Conclusion.}
\label{conclusion}

We have performed a RMT analysis of the quantum four-state Potts chain
for different sizes of the quantum chain, and for different unfolding
methods. Our calculations unambiguously exhibit a GOE statistics and
excludes GUE (and GSE) statistics.  

Our results indicate that there exists, for arbitrary size \( \, L \),
a symmetric unitary matrix \( \, K \), such that \( \, K\cdot H^{t}\,
=\, H\cdot K \).  This can be checked exactly for small lattice sizes
\( L\, =\, 2,3, 4, 5,6 \).  We conjecture that such a relation exists
for all sizes of the chain, and for each of the $4 L$ blocks
(\ref{HKredu2}). The existence of \( \, K \) would account for the
statistics we find (GOE rather GUE).

When the hamiltonian becomes integrable our analysis shows the change
from the (generic) GOE distribution to a Poisson distribution and this
reduction does not require the spectral curve to be of genus $\, 0$ or
$\, 1$.

It is thus interesting to combine this RMT approach with more
algebraic methods developed in previous
publications~\cite{BeMaVi92,Ma86,22}. These methods will give the
algebraic subvarieties where a ``higher genus'' integrability may
appear, if any ... (the infinite set of these algebraic subvarieties
can be obtained exactly for the four-state chiral Potts
model~\cite{Prepa}).  As we have shown, the change in the level spacing
statistics will signal integrability, bypassing the difficulties of
the analytical approach.

\vskip .3cm 

{\bf Acknowledgments :} We would like to thank 
R.J. Baxter and J.H.H. Perk. for
 illuminating discussions on the four-state chiral Potts model.  We also
thank S. Boukraa and R. Attal for careful readings of the manuscript. 

\appendix

\section{Appendix}

\subsection{Generalized time-reversal invariance.}
\label{timerev2}

The anti-unitary time-reversal operator $\, T$ 
can be expressed as the product
of a unitary operator K and the conjugation operator C, namely
$\, T\, =\, K\cdot C$,
 where $\, T$ is projectively an involution, namely 
\( T^{2}=\lambda .\textrm{Id} \), and
 where $\, \textrm{Id}$ denotes the
identity operator. The factor \( \lambda  \)
being equal to \( \pm 1 \) as a consequence of the unitarity of \( K \).
The time-reversal operator \( T \) must change the time evolution
operator according to:\begin{equation}
\label{deftimevol}
T\, e^{-iHt}\, T^{-1}\, =\, e^{+iHt}
\end{equation}
or equivalently\begin{equation}
\label{deftimevol2}
T\, e^{-iHt}\, T\, =\, \lambda \cdot e^{+iHt}
\end{equation}

Expanding  (\ref{deftimevol}), or  (\ref{deftimevol2}), in the time
variable \( t \),
one gets for every order \( n \):
\[
K\left( H^{\star }\right) ^{n}K^{\star }\, 
=\, \lambda \cdot  H^{n}\qquad \forall
\, n\]
 yielding only two equations:
\begin{eqnarray}
KK^{\star } & = & \lambda \label{TimeRev0} \\
KH^{\star }K^{\star } & = & \lambda \cdot H \label{TimeRev1} 
\end{eqnarray}
For an hermitian hamiltonian,  (\ref{TimeRev1}) becomes using
 (\ref{TimeRev0}) : 
\begin{equation}
\label{HR3}
H\, =\, K\cdot H^{t}\cdot K^{-1}
\end{equation}
 where \( \, H^{t}\,  \) is the transpose of H. Since the operator \( K \)
is a unitary one,  (\ref{TimeRev0}) yields :
\begin{equation}
\label{SAS}
K\, =\, \lambda .\left( K^{\star }\right) ^{-1}\, =\, \lambda .K^{t}
\end{equation}
where \( K^{t} \) denotes the transpose of  \( K \). Recalling 
that \( \lambda =\pm 1 \),
we see, from (\ref{SAS}), that \( K \) {\em must be either a symmetric
 or an antisymmetric unitary
matrix}.\\

\subsection{Matrix $\, K$ 
 as  block diagonal matrices for $\, L\, = \, 3$ for hamiltonian (\ref{hamilt2}).}
\label{asablock}

Let $\, \omega\, $ be the third root of unity :
$ \,\omega\,  = \,-1/2 - \,i \sqrt {3}/2$, and the  consider $\, L\, = \, 3$ case with  $\, \alpha \, = \, 0, \, 1, \, 2, \, \cdots , \, 11$ indexing
the twelve blocks defined in (\ref{bloblo}).

For  hamiltonian (\ref{hamilt2})
and 
  $\, K_{block} \, = \,P_{ZXX} \cdot K_{ZXX} \cdot P_{ZXX}^{t} $ 
in (\ref{HKredu}), the  off-diagonal blocks of  
$ K_{block}$  vanish and  one finds
 the following expressions for  the  diagonal blocks 
$\, k_{\alpha, \alpha}$ : 
\begin{eqnarray}
\label{k117}
k_{3,3}\, = \, \, 
\left [\begin {array}{cccccc}
 1&0&0&0&0&0\\
\noalign{\medskip}0&0&1&0&0&0\\
\noalign{\medskip}0&1&0&0&0&0\\
\noalign{\medskip}0&0&0&1&0&0\\
\noalign{\medskip}0&0&0&0&1&0\\
\noalign{\medskip}0&0&0&0&0&1
\end {array}\right ], \quad \quad \quad 
k_{6,6}\, = \, \, 
\left [\begin {array}{cccccc} 
1&0&0&0&0&0\\
\noalign{\medskip}0&1&0&0&0&0\\
\noalign{\medskip}0&0&1&0&0&0\\
\noalign{\medskip}0&0&0&0&1&0\\
\noalign{\medskip}0&0&0&1&0&0\\
\noalign{\medskip}0&0&0&0&0&1
\end {array}\right ] \nonumber
\end{eqnarray}
\begin{eqnarray}
\label{k1171}
k_{1,1}\, = \, \, 
\left [\begin {array}{ccccc} 
0&0&1&0&0\\
\noalign{\medskip}0&1&0&0&0\\
\noalign{\medskip}1&0&0&0&0\\
\noalign{\medskip}0&0&0&\omega&0\\
\noalign{\medskip}0&0&0&0&1
\end {array}\right ], \quad \quad \quad \quad 
k_{4,4}\, = \, \, 
\left [\begin {array}{ccccc} 
\omega&0&0&0&0\\
\noalign{\medskip}0&0&1&0&0\\
\noalign{\medskip}0&1&0&0&0\\
\noalign{\medskip}0&0&0&\omega&0\\
\noalign{\medskip}0&0&0&0&1
\end {array}\right ]
, \quad  \nonumber
\end{eqnarray}
\begin{eqnarray}
\label{k1172}
k_{7,7}\, = \, \, 
\left [\begin {array}{ccccc}
 \omega&0&0&0&0\\
\noalign{\medskip}0&1&0&0&0\\
\noalign{\medskip}0&0&1&0&0\\
\noalign{\medskip}0&0&0&0&1\\
\noalign{\medskip}0&0&0&1&0
\end {array}\right ], \quad \quad \quad \quad 
k_{10,10}\, = \, \, 
\left [\begin {array}{ccccc} 
\omega&0&0&0&0\\
\noalign{\medskip}0&0&1&0&0\\
\noalign{\medskip}0&1&0&0&0\\
\noalign{\medskip}0&0&0&1&0\\
\noalign{\medskip}0&0&0&0&\omega
\end {array}\right ] \nonumber
\end{eqnarray}
and $\,  k_{9,9}\, = \, \, k_{3,3}$, the block 
$\, k_{0,0}\, $ {\em is the same as the one for hamiltonian} (\ref{hamilt}). 
Furthermore $\,  k_{2,2}\,$ is equal to block $\,  k_{1,1}\,$
where $ \, \omega $ is changed into  $ \, \omega^2 $, i.e.
$\,  k_{2,2}(\omega)\,= \, k_{1,1}(\omega^2)$.
Similarly  one gets $\,  k_{5,5}(\omega)\,=  k_{4,4}(\omega^2)\,$
as well as $\,  k_{8,8}(\omega)\,= \,  k_{7,7}(\omega^2)\,$
and  $\,  k_{11,11}(\omega)\,=\,  k_{10,10}(\omega^2)$.


\vskip .2cm

\vskip 1.2cm


\begin{thebibliography}{10}

\bibitem{Rosen}N. Rosenzweig and C. E. Porter, Phys. Rev \textbf{120}, 1698
(1960).

\bibitem{Gu}M.C. Gutzwiller, \emph{Chaos in Classical and Quantum
Mechanics}, Interdisciplinary Applied Mathematics, Springer-Verlag,
New York Berlin Heidelberg, 1990. 

\bibitem{Wig55}E.P. Wigner, Ann. Math. \textbf{62}, 
548 (1955); \textbf{65}, 203
(1957); \textbf{67}, 325 (1958) 

\bibitem{key-8}F. J. Dyson, J. Math. Phys. 
\textbf{3}, 140 (1962). F. J. Dyson, J. Math. Phys.
\textbf{3}, 157 (1962), F. J. Dyson, J. Math. Phys. \textbf{3}, 166 (1962).

\bibitem{Meh91}M.L. Mehta, \emph{Random Matrices}, 2nd ed. (Academic
Press, San Diego,1991).

\bibitem{FaWu70} C. Fan and F.Y.
Wu, Phys. Rev. \textbf{B 3}, 723 (1970). 

\bibitem{Fel73c} B.U. Felderhof,
Physica \textbf{66}, 509 (1973). 

\bibitem{Bohi} O. Bohigas, \emph{Les Houches, 1989, Chaos et physique
quantique}, M-J. Giannoni et al editor, North Holland (1991).

\bibitem{rmttransfer}H. Meyer, J.-C. Anglès d'Auriac, 
and J-M. Maillard, Phys. Rev. \textbf{E 55}, 5380 (1997). 

\bibitem{criterion} M. T. Jaekel and J. M. Maillard,
J. Phys.\textbf{A18}, 1229 (1985)

\bibitem{GeHaLeMa87}A. Georges, D. Hansel, P. Le Doussal and J-M. Maillard
J. Phys.\textbf{A20}, 5299 (1987)

\bibitem{Kast}P.W. Kasteleyn, \emph{ Exactly Solvable Lattice Models
}, in \emph{ Proc. of the 1974 
Wageningen Summer School: Fundamental
 problems in statistical mechanics III
}, edited by E.G.D. Cohen
 (North--Holland, Amsterdam, 1974) pp. 103--155. 

\bibitem{Sut70}B. Sutherland, J. Math. Phys. \textbf{11}, 3183 (1970). 

\bibitem{Gaudin} M. Gaudin, \emph{La fonction d'onde de Bethe},
Collection du C.E.A. S\'erie Scientifique"
Masson, Paris (1983)

\bibitem{LiWu72}E.H. Lieb 
and F.Y. Wu, in \emph{
Phase Transitions and Critical Phenomena}, edited by C. Domb and
M. Green (Academic Press, New York, 1972), Vol. 1,
pp. 331--49 

\bibitem{Bax82}R. Baxter, \emph{Exactly Solved Models in Statistical Mechanics} (Academic
Press, New York, 1982). 

\bibitem{BaPeAu88}R.J. Baxter, J.H.H. Perk and H. Au-Yang, 
Phys. Lett.  \textbf{ A 128}, 138 (1988)

\bibitem{chiral}H. Au-Yang, B.M.McCoy, J.H.H.Perk, S.Tang and
M-L. Yan, Phys. Lett.  \textbf{A 123}, 219 (1987)

\bibitem{Howes}S. Howes, L. P. Kadanoff  and M. den Nijs, Nucl. Phys. \textbf{B 215}, 169
(1983). 

\bibitem{vanG}G. von Gehlen and V. Rittenberg, Nucl. Phys. \textbf{B 257}, 351
(1985). 

\bibitem{Dolan}L. Dolan and M. Grady, Phys. Rev.  \textbf{D 25}, 1587
(1982). 

\bibitem{DaDeKlaMcCoMe93} S. Dasmahapatra,
R. Dedem, T. R. Klassen, B. M. McCoy,
and E. Melzer, {\em Quasi-particles, Conformal Field Theory and
$q$-series} in \emph{  Yang-Baxter
Equations in Paris}, edited by J.-M. Maillard (World Scientific,
Singapore, 1993). 

\bibitem{Albert} G. Albertini, {\em
Exact Spectrum of the 3-state Potts Chain}, in {\em
Yang-Baxter Equations in Paris}, edited by J-M. Maillard
 (World Scientific, 1993) 

\bibitem{BeMaVi92} M. Bellon, J.-M. Maillard, and
C.-M. Viallet, Phys. Lett. \textbf{B 281}, 315 (1992). 


\bibitem{BrAdA96}H. Bruus and J.-C. Anglès d'Auriac, Europhys. Lett. 
\textbf{35}, 321 (1996). 

\bibitem{HsAdA93}T.C. Hsu and J.-C. Anglès d'Auriac, Phys. Rev. \textbf{B}
 \textbf{47}, 14291 (1993). 

\bibitem{hm4}H. Meyer, J.-C. Anglès d'Auriac  and H. Bruus, J. Phys. 
\textbf{A}:Math. Gen. \textbf{L483} (1996). 

\bibitem{hmth}H. Meyer, Ph.D. thesis, Univ. J. Fourier, Grenoble, France, 1996.

\bibitem{DupuisMont} N. Dupuis and G. Montambaux, {\em Aharonov-Bohm
flux and statistics of energy levels in metals}, Phys. Rev. \textbf{B
43}, 14390 (1991)

\bibitem{PoZiBeMiMo93}D. Poilblanc, T. Ziman, J. Bellisard, F. Mila  and
G. Montambaux, Europhys. Lett. \textbf{22}, 537 (1993). 

\bibitem{Rob}M. Robnik and M. V. Berry, J. Phys. \textbf{A 19}, 669 (1986).

\bibitem{JMMS80} M. Jimbo, T Miwa, Y. Mori and M. Sato,
Physica {\bf D 1} 80-158 (1980).          

\bibitem{TraWi93} C.A. Tracy and H. Widom,  Com. Math. Phys. {\bf 163}
(1994), 33-72

\bibitem{FoWi02a} P.J. Forrester and N.S. Witte ``Application of the
$\tau$-function theory of Painlev\'e equations to random matrices:
$P_V$, $P_{III}$, the LUE, JUE and CUE'' arXiv:math-ph/0201051

\bibitem{FoWi02b} P.J. Forrester and N.S. Witte ``$\tau$-function
evaluation of gap probabilities in orthogonal and symplectic matrix
ensembles'' arXiv:math-ph/0203049

\bibitem{FoWi02c} P.J. Forrester and N.S. Witte ``Application of the
$\tau$-function theory of Painlev\'e equations to random matrices:
$P_{VI}$, the JUE, CyUE, cJUE and scaled limits''
arXiv:math-ph/0204008

\bibitem{MoPoBeSi93}G. Montambaux, D. Poilblanc, J. Bellisard  and C. Sire,
 Phys. Rev.  Lett. \textbf{70}, 497 (1993). 

\bibitem{vEGa94}P. van Ede van der Pals and P. Gaspard, Phys. Rev.
\textbf{E 49}, 79 (1994). 

\bibitem{BrAdA97}H. Bruus and J.-C. Angl\`es d'Auriac, cond-mat/9610142 (1996).
 H. Bruus and J.-C. Angl\`es d'Auriac, Phys. Rev. \textbf{B  55}, 9142 (1997). 

\bibitem{Pok}Y. A. Bashilov and S. V. Pokrovsky, Comm. Math. Phys. 
 \textbf{76}, 129 (1980). 

\bibitem{Marcu}M. Marcu, A. Regev and V. Rittenberg, J. Math. Phys.
 \textbf{22}, 2740 (1981). 

\bibitem{seldu}
J. C. Angl\`es d'Auriac and Ferenc Igl\'oi.
Phys. Rev. {\bf E 58}, 241 (1998).

\bibitem{lapack} lapack library: http://www.netlib.org/lapack/

\bibitem{wigner}E.P. Wigner, \emph{Group Theory and its application to
the quantum mechanics of atomic spectra},  (Academic Press,
New-York and London, 1959). 


\bibitem{key-2}H. Weyl, \emph{Classical Groups,} Princeton Univ. Press,
 Princeton New Jersey 1946.

\bibitem{key-3}C. Chevalley, \emph{Theory of Lie Groups,} pp 16-24,
Princeton Univ.  Press, Princeton New Jersey 1946.

\bibitem{key-4}Jean-Alexandre Dieudonné, \emph{La Géométrie des
Groupes Classiques}, Ergeb. d. Math. vol 5 Springer Berlin 1955.

\bibitem{vonG5371}G. von Gehlen, J. Phys. \textbf{A 24}, 5371 (1991).

\bibitem{nonherm}G. von Gehlen, Int. Journ. Mod. Phys. \textbf{B 8},
3507 (1994).

\bibitem{Prepa} J-C. Angl\`es d'Auriac, J-M. Maillard and
C. M. Viallet, in preparation.

\bibitem{web} 
http://crtbt.polycnrs-gre.fr/theo/pagesperso/dauriac/QPOTTS/QPotts.html

\bibitem{Ma86} J-M. Maillard, J. Math. Phys. \textbf{27}, 2776, (1986)

\bibitem{22} M.P. Bellon,
J.-M. Maillard, and C.-M. Viallet, Phys. Lett. \textbf{ B 260}, 87 (1991). 

\bibitem{JaMa85} M. T. Jaekel and
J.-M. Maillard, J. Phys. \textbf{ A 18}, 1229 (1985). 

\bibitem{MaRo94} 
J.-M. Maillard and G. Rollet, J. Phys. \textbf{A 27}, 6963 (1994). 

\bibitem{MeAnMaRo94}H.Meyer, J-C. Angl\`es d'Auriac,
 J-M. Maillard and G. Rollet,  Physica. \textbf{ A 208}, 223 (1994). 

\bibitem{HaMa88} D. Hansel and J-M. Maillard, Phys. Lett. \textbf{A
133}, 11, (1988)

\bibitem{MaRa83} J.-M. Maillard and R. Rammal, J.  Phys.  \textbf{ A
16}, 353 (1983).

\bibitem{Lets} S.Boukraa and J-M. Maillard,
J. Stat. Phys. \textbf{102}, 641, (2001)

\end{thebibliography}
\end{document}